\documentclass[12pt]{article}
\let\ssection=\section
\renewcommand{\section}{\setcounter{equation}{0}\ssection}

\newcommand\mathC{\mkern1mu\raise2.2pt\hbox{$\scriptscriptstyle|$}
        {\mkern-7mu\rm C}} 
\newcommand{\mathR}{{\rm I\! R}}         

\newcommand\bi{\begin{itemize}}
\newcommand\ei{\end{itemize}}
\newcommand\be{\begin{equation}}
\newcommand\ee{\end{equation}}

\newcommand{\dd}{\ensuremath{{\delta}}}
\newcommand{\al}{\ensuremath{{\alpha}}}

\newcommand{\pl}{\ensuremath{{\partial}}}

\begin{document}
\begin{titlepage}

\begin{center}
{\large\bf On Hamilton-Jacobi Theory as a Classical Root of Quantum Theory}
\end{center}

\vspace{0.8 truecm}

\begin{center}
        J.~Butterfield\footnote{email: jb56@cus.cam.ac.uk;
            jeremy.butterfield@all-souls.oxford.ac.uk}\\[10pt] All Souls 
College\\ Oxford OX1 4AL
\end{center}

\begin{center}
        27 February 2003
\end{center}

\vspace{0.4 truecm}

\begin{abstract}
This paper gives a technically elementary treatment of some aspects of  
Hamilton-Jacobi theory, especially in relation to the calculus of variations. 
The second half of the paper describes the application to geometric optics, the 
optico-mechanical analogy and the transition to quantum mechanics. Finally, I 
report recent work of Holland providing a Hamiltonian formulation of the 
pilot-wave theory.   
 \end{abstract}
\vspace{0.5 truecm}

\noindent Forthcoming in Elitzur, A.C., Dolev, S., \& Kolenda, N. (editors), 
{\em Quo Vadis
Quantum Mechanics?  Possible Developments in Quantum Theory in the 21st
Century}, New York: Springer, 2004. (The Frontier Series: Monographs and Books 
on Frontiers of
Modern Physics)

  \vspace{0.5 truecm}

\noindent{\em `Dont worry, young man: in mathematics, none of us really 
understands any idea---we just get used to them.'}\\
John von Neumann, after explaining (no doubt very quickly!) the method of 
characteristics (i.e. Hamilton-Jacobi theory) to a young physicist, as a way to 
solve his problem; to which the physicist had replied `Thank you very much; but 
I'm afraid I still don't understand this method.'

\end{titlepage}

\section{Introduction}\label{sec:intro}
In the eighty years since its discovery in the mid-1920s, quantum mechanics has 
gone from strength to strength. It has repeatedly been proved successful, to a 
high degree of accuracy, in domains of application very different from its 
original one. For example, although it was devised for systems of atomic 
dimensions ($10^{-8}\mbox{cm}$), it has since proven accurate for scales much 
smaller (cf. the nuclear radius of ca.
$10^{-12}\mbox{cm}$) and vastly larger (cf. superconductivity and
superfluidity, involving scales up to $10^{-1}\mbox{cm}$). Similarly, if we 
think of domains of application, not as length (or energy) scales, but as types 
of ``stuff'' to which the theory applies. Though quantum mechanics was first 
devised to apply to matter (i.e. electrons and protons, the more ``obvious'' 
constituents of atoms), it was soon extended to fields, i.e. the 
electromagnetic field: indeed, matter soon became regarded as excitations in 
associated fields. Similarly, if we think of domains of application as types of 
force: though first devised for electromagnetic forces, quantum mechanics now 
successfully describes the weak and strong forces. Indeed, similarly for 
`domains' understood naively, as regions of the universe: quantum mechanics has 
also been applied with great success to astronomy---the obvious examples being 
the use of nuclear physics in theories of stellar structure and
evolution, and of particle physics in theories of the early universe.

So quantum mechanics has been an amazing success story. I stress this point at 
the outset, for two reasons. First: it is, unfortunately, all too easy to get 
used to success.  Nowadays, both physicists, for whom the various quantum 
theories have become everyday professional 
tools, and the wider scientifically literate public, can easily lose their 
sense of wonder at this immense  success. So it is worth remembering how 
contingent, and surprising, it is.

My second reason is more specific to work in the foundations and-or philosophy 
of quantum theory. This work focusses on the interpretative problems, 
especially the measurement problem, that still confront quantum mechanics, 
despite its immense empirical success: hence this volume's question `{\em Quo 
vadis}, quantum mechanics?' Of course, I endorse that focus: it is crucially 
important to address these problems. But in addressing them, it is salutary to 
recall this success, as an intellectual backdrop. Indeed, not only is it 
salutary: it might also  be heuristically useful---though of course,  different 
researchers, with their different intellectual temperaments, will take this 
success to give different heuristic clues about `{\em Quo vadis}, quantum 
mechanics?'.  For example, an Everettian philosopher such as Saunders (??this 
volume) may see the success of the established quantum theoretic formalisms as 
supporting their position: certainly, heterodox quantum theories such as 
dynamical models of wave-function collapse face  an enormous task in recovering 
that success. On the other hand, a theoretical physicist who is searching for a 
successor to quantum mechanics---whether to solve these interpretative problems 
or to reconcile the quantum with general relativity's treatment of gravitation, 
or both (such as 't Hooft, ??this volume)---may scrutinize the details of this 
empirical  success for clues about how present-day quantum mechanics might be 
an effective, i.e. phenomenological, theory. As 't Hooft wittily puts it: we 
can ask, not '{\em Quo vadis}, quantum mechanics?', but rather `{\em Unde 
venis?}'---`Where do you come from?'  

This paper will likewise ask `{\em Unde venis}, quantum mechanics?': though I 
humbly admit that I will interpret this question in a  retrospective and 
expository sense, rather than in 't Hooft's wonderfully forward-looking and 
creative sense. To be specific: I propose to discuss Hamilton-Jacobi theory as 
a classical root of quantum mechanics.

One part of this story is well known to physicists and philosophers and 
historians of physics. Namely: Hamilton-Jacobi theory as a method of 
integrating Hamilton's equations (using Jacobi's theorem, action-angle 
variables etc.), and the use made of this integration theory in nineteenth 
century celestial mechanics, and thereby in the old quantum theory.

There is however another part of this story that seems much less known by this 
community: viz. Hamilton-Jacobi theory  understood from the perspective of the 
calculus of variations (as worked out by such masters as Hilbert and 
Carath\'{e}odory), and how this understanding motivates deBroglie's and 
Schr\"{o}dinger's proposal to extend Hamilton's optico-mechanical analogy, thus 
creating quantum mechanics (as wave mechanics). So I propose to present this 
part of the story: or rather, since this part could fill a book---selected 
pieces of it! (My (2003, 2003a) discuss some other, philosophical, aspects.) At  
the end of the paper, I shall also briefly return to `{\em Quo vadis?}', i.e. 
to a current interest in the foundations of quantum theory: viz. the pilot-wave 
theory---on which Hamilton-Jacobi theory casts some light. But I begin, in the 
next Subsection, with a more detailed prospectus.

\subsection{Introducing Hamilton-Jacobi theory}\label{ssec;introHJ}
Hamilton-Jacobi theory is a general theory, rich in analytic and geometric 
ideas, that unifies three apparently disparate topics: systems of first order 
ordinary differential equations, first order partial differential equations, 
and the calculus of variations. Roughly speaking, Hamilton-Jacobi theory shows 
that the following problems are  equivalent:---\\
\indent (ODE): solving a canonical system of first order ordinary differential 
equations ($2n$ equations for $2n$ functions of a parameter $t$ in which all 
variables' first derivatives are given by partial derivatives of one and the 
same function); e.g. Hamilton's equations in Hamiltonian mechanics.\\
\indent (PDE): solving a first order partial differential equation in which the 
unknown function does not occur explicitly; e.g. the Hamilton-Jacobi equation 
in mechanics.\\
\indent (CV): solving the ``basic'' calculus of variations problem of finding 
$n$ functions $q_1,\dots,q_n$ of a parameter $t$ that make stationary a 
line-integral of the form $\int L(q_i,{\dot q_i},t) \; dt$, where the dot 
denotes differentiation with respect to $t$; e.g. Hamilton's principle in 
Lagrangian mechanics, or Fermat's principle in geometric optics.

A bit more precisely: elementary Lagrangian and Hamiltonian mechanics show 
(ODE) and (CV) to be equivalent for the case of fixed end-points. 
Hamilton-Jacobi theory extends this equivalence by considering, not a single 
solution of the canonical equations (a single line-integral) but a whole field 
of solutions, i.e. line-integrals along all the curves  of a space-filling 
congruence (so that the end-points lie on hypersurfaces transverse to the 
congruence). The initial conditions of a problem then become the specification 
of a function's values on such a hypersurface, instead of an initial 
configuration and momentum (or an initial and final configuration): hence the 
occurrence of partial differential equations.  

The main aim of this paper is to explain (in an elementary way) these 
equivalences and some related results. This explanation will later (Sections 
\ref{sec:optics} and \ref{sec:mechs}) provide us with a perspective on the 
optico-mechanical analogy and quantum mechanics (specifically, wave mechanics). 
But there is also a pedagogic rationale for presenting these results. Most 
physicists learn Hamilton-Jacobi theory only as part of analytical mechanics; 
and almost all the mechanics textbooks present, in addition to the equivalence 
of (ODE) and (CV) for fixed end-points, only the use of Hamilton-Jacobi theory 
as a method of integrating Hamilton's equations---indeed rendering the 
integration trivial. The central result here is Jacobi's theorem: that given a 
complete integral of the Hamilton-Jacobi equation (typically found by 
separation of variables), one can obtain solutions of Hamilton's equations just 
by differentiation.  This is a remarkable result, which lies at the centre of a 
beautiful geometric theory of the integration of first order partial 
differential equations: a theory which reduces the integration problem to that 
of integrating a suitable system of ordinary differential equations (the 
characteristic equations). But almost all the mechanics textbooks present 
Jacobi's theorem using just canonical transformation theory: as a result, they 
do not describe this general  integration theory---and more generally, they do 
not show the role of  geometric ideas, nor of the calculus of variations with 
variable end-points. 

This textbook tradition is of course understandable. Textbooks must emphasise 
problem-solving; and the use of a complete integral of the Hamilton-Jacobi 
equation to solve Hamilton's equations, is crucially important, for several 
reasons. As to problem-solving, it is `the most powerful method known for exact 
integration, and many problems which were solved by Jacobi cannot be solved by 
other means' (Arnold 1989, p. 261). Besides, it is conceptually important: it 
leads on to action-angle variables, which are central both to classical 
mechanics (e.g. in the Liouville-Arnold theorem, and in perturbation theory) 
and the old quantum theory. 

But though understandable, this tradition is also regrettable. For the result 
is that
most physicists understand well only the equivalence of (ODE) and (CV) for 
fixed end-points, and a part of the equivalence of (PDE) and (ODE)---the part  
expressed by Jacobi's theorem. Besides, they understand these matters only in 
the context of mechanics. This is a pity, for two reasons.

First, it is worth stressing that all these equivalences and related other 
results, are purely mathematical and so entirely general. Second, the 
equivalences and results that get omitted from most mechanics textbooks are at 
least as rich as those included; in particular, in their use of geometric 
ideas. I might add: `in their use of optical ideas'. Indeed, Hamilton  
developed his work in mechanics in deliberate analogy with his previous work in 
optics.\footnote{For a glimpse of the history, which I will not discuss, cf. 
e.g.: for mechanics, Dugas (1988), Whittaker (1959); for optics, Whittaker 
(1952), Buchwald (1989); and for mathematics: Kline (1970, Chap. 30).} And as 
we shall see: both Fermat's principle (roughly, that a light ray travels the 
path that takes least time) and  Huygens's principle (roughly, that given a 
wave-front, a later wave-front is the envelope of spherical waves spreading 
from the points of the given wave-front) stand at the centre of Hamilton-Jacobi 
theory. They involve each of the above mathematical problems, in optical guise: 
viz. the description of light in terms of rays (exemplifying (ODE)), in terms 
of wavefronts (cf. (PDE)), and by means of variational principles (cf. (CV)).

Accordingly, I propose to expound some of these equivalences and connections, 
as mathematics (Sections \ref{sec:CVtoHJ} to \ref{sec:fopdes}). Then I will 
illustrate them with geometric optics and the optico-mechanical analogy 
(Sections \ref{sec:optics} and \ref{sec:mechs}).

To be both brief and elementary, this exposition must be {\em very} selective. 
In particular, I will say nothing about: (i) weak solutions; (ii) the use of 
phase space; (iii) issues about the global existence of solutions, including 
focussing and caustics.\footnote{A few pedagogic references: for (i) Logan 
(1994, Chap. 3), Stakgold (1967); for (ii), Arnold (1989, Chap.s 8, 9), 
Littlejohn (1992), Taylor (1996, Section 1.15); for (iii), Arnold (1989, 
Appendices 11, 16), Benton (1977), Taylor (1996, Section 6.7). Of these topics, 
(ii) and (iii) are closest to this paper's interests in geometry, and in the 
transition between classical and quantum mechanics. For (ii), i.e. 
Hamilton-Jacobi theory in phase space, beautifully illustrates symplectic 
geometry; and (ii) and (iii) are  crucial in both quantization theory and 
semiclassical mechanics.} Another omitted topic lies closer to our concerns: I 
will {\em not} present the theory surrounding Jacobi's theorem, i.e. 
Hamilton-Jacobi theory as an integration theory for first order partial 
differential equations.   For though I have complained that this is absent from 
the mechanics books, it {\em is} in some books on mathematical 
methods.\footnote{Especially Courant and Hilbert (1962, Chap. II.1-8); cf. also 
e.g. Webster (1950, Chap 2) and John  (1971, Chap 1). In order to be 
elementary, I will also avoid all use of modern differential geometry, 
including even the distinction between contravariant and covariant indices. 
Though modern geometry has transformed our understanding of differential 
equations and the calculus of variations (and the sciences of mechanics and 
optics), I shall only need the intuitive geometry familiar from multivariable 
calculus.}

Instead, I will adopt an approach that emphasises the calculus of variations. 
The main ideas here seem to be due to Carath\'{e}odory and Hilbert. Here again, 
I must be selective: I will simply pick out within this approach, one line of 
thought, found for example in the first half of Rund (1966). (Rund proves some 
results which I will only state; and he cites the original papers.) Though 
selective, this exposition will give a good sense of the triangle of 
equivalences between (ODE), (PDE) and (CV); indeed, we will get such a sense 
already by the end of Section \ref{sec:CELEfields}. Sections \ref{sec:HbtIntgl} 
to \ref{sec:fopdes} will add to this a discussion of three topics, each leading 
to the next. They are, respectively: Hilbert's independent integral; treating 
the integration variable of the variational problem on the same footing as the 
other coordinates; and integration theory.

 Thereafter, Sections \ref{sec:optics} et seq. return us to physics. Section 
\ref{sec:optics} discusses geometric optics; and Section \ref{sec:mechs}, the 
optico-mechanical analogy and wave mechanics. Section \ref{sec:mechs} also 
leads us back to the foundations of quantum mechanics: which I take up briefly 
in (the last) Section \ref{sec:pwthy}. Here I will call attention to the role 
of Hamilton-Jacobi theory in the pilot-wave theory of deBroglie and Bohm; and 
more specifically advertise Holland's recent work (2001, 2001a), which provides 
a Hamiltonian formulation of the pilot-wave theory. 

\section{From the calculus of variations to the Hamilton-Jacobi 
equation}\label{sec:CVtoHJ}
\subsection{The calculus of variations reviewed}
\label{ssec;cvreview}
We begin by briefly reviewing the simplest problem of the calculus of 
variations; with which we will be concerned throughout the paper. This is the 
variational problem (in a notation suggestive of mechanics)
\be
\dd I := \dd I[q_i] = \dd \int^{t_1}_{t_0} L(q_i,{\dot q}_i,t) dt = 0 \;\; , 
\;\;\; i = 1,\dots,n
\label{actionwithpara=t}
\ee
where $[\;]$ indicates that $I$ is a functional, the dot denotes 
differentiation with respect to $t$, and 
$L$ is to be a $C^2$ (twice continuously differentiable) function in all $2n+1$ 
arguments. $L$ is the {\em Lagrangian} or {\em fundamental function}; and $\int 
L \; dt$ is the {\em fundamental integral}. We will discuss this only locally; 
i.e. we will consider a fixed simply connected region $G$ of a 
$(n+1)$-dimensional real space $\mathR^{n+1}$, on which there are coordinates 
$(q_1,\dots,q_n,t) =: (q_i,t) =: (q,t)$.

The singling out of a coordinate $t$ (called the {\em parameter} of the 
problem), to give a parametric representation of curves $q(t) := q_i(t)$, is 
partly a matter of notational clarity. But it is of course  suggestive of the 
application to mechanics, where  $t$ is time, $q$ represents the system's 
configuration and $(q_i,t)$-space is often called `extended configuration  
space' or `event space'. Besides, the singling out of $t$  reflects the fact 
that though it is usual to assume that $L$ (and so the fundamental integral) is 
invariant under arbitrary transformations (with non-vanishing Jacobian) of the 
$q_i$, we do not require the fundamental integral to be independent of the 
choice of $t$. Indeed we shall see (at the end of this Subsection and in 
Section \ref{sec:paraadditional}) that allowing this dependence is necessary 
for making Legendre transformations.\footnote{Of course, the calculus of 
variations, and Hamilton-Jacobi theory, {\em can} be developed on the 
assumption that the fundamental integral is to be parameter-independent---if it 
could not be, so much the worse for relativistic theories! But the details, in 
particular of how to set up a canonical formalism, are different from what 
follows. For these details, cf. e.g. Rund (1966, Chapter 3).}

A necessary condition for $I$ to be stationary at the $C^2$ curve $q(t) := 
q_i(t)$---i.e. for $\dd I = 0$ in comparison with other $C^2$ curves that  (i) 
share with $q(t)$ the end-points $q(t_0), q(t_1)$ and (ii) are close to $q(t)$ 
in both value and derivative throughout $t_0 < t < t_1$---is that: $q(t)$ 
satisfies for $t_0 < t < t_1$ the $n$ second-order Euler-Lagrange (also known 
as: Euler, or as Lagrange!) equations
\be
\frac{d}{dt}L_{{\dot q}_i} - L_{q_i} = 0 \;\;\;\; i = 1,\dots,n.
\label{elpara=t}
\ee
A curve satisfying these equations is called an {\em extremal}.
 
We will not need to linger on the usual derivation of these equations: we will 
later see them derived without using a single fixed pair of end-points. Nor 
need we linger on several related matters taken up in the calculus of 
variations, such as: the distinction between stationarity and extrema (i.e. 
maxima or minima), in particular the conditions for a curve to be an extremum 
not just a stationary point (e.g. conditions concerning the second variation of 
the fundamental integral, or Weierstrass' excess function); the distinction 
between weak and strong stationary points and extrema; and the use of weaker 
assumptions about the smoothness of the solution and comparison curves.
 
But it is important to consider the canonical form of our variational problem. 
In physics, the most frequent example of this is the expression of Hamilton's 
principle within Hamiltonian mechanics; i.e. Hamilton's principle with the 
integrand a function of both $q$s and $p$s, which are to be varied 
independently. But the correspondence between the Lagrangian form of the 
variational problem (above) and the canonical form is general (purely 
mathematical). 

Thus, under certain conditions the variational problem eq. 
\ref{actionwithpara=t} has an equivalent form, whose Euler-Lagrange equations 
are $2n$ first order equations. To this end, we introduce ``momenta'' 
\be
p_i := L_{{\dot q}_i} \; ;
\label{definepsubi}
\ee 
and (recalling that $L$ is $C^2$) we assume that the Hessian with respect to 
the ${\dot q}$s does not vanish in the domain $G$ considered, i.e. the 
determinant 
\be
\mid L_{{\dot q}_i{\dot q}_j}\mid \neq 0 \; ;
\label{nonzerohessian}
\ee
so that eq. \ref{definepsubi} can be solved for the ${\dot q}_i$ as functions 
of $q_i, p_i,t$. 

Then the equations
\be
p_i = L_{{\dot q}_i} \;\;\; {\dot q}_i = H_{p_i} \;\;\;
L(q_i,{\dot q}_i,t) + H(q_i,p_i,t) = \Sigma_i {\dot q}_ip_i
\label{legtrfmnsymmicform2nd}
\ee 
represent a {\em Legendre transformation} and its inverse; where in the third 
equation ${\dot q}_i$ are understood as functions of $q_i,p_i,t$ according to 
the inversion of eq. \ref{definepsubi}. The function $H(q_i,p_i,t)$ is called 
the {\em Legendre ({\rm or:} Hamiltonian) function} of the variational problem, 
and the $q$s and $p$s are called {\em canonically conjugate}. It follows that 
$H$ is $C^2$ in all its arguments, $H_t = -L_t$, and $\mid L_{{\dot q}_i{\dot 
q}_j}\mid = \mid H_{{p}_i{p}_j}\mid^{-1}$. Besides, any  $H(q_i,p_i,t)$ that is 
$C^2$ in all its arguments, and has a non-vanishing Hessian with respect to the 
$p$s, $\mid H_{{p}_i{p}_j}\mid \neq 0$, is the Legendre function of a  $C^2$ 
Lagrangian $L$ given in terms of $H$ by eq. \ref{legtrfmnsymmicform2nd}. 

Applying this Legendre transformation, the Euler-Lagrange equations eq. 
\ref{elpara=t} go over to the {\em canonical system}
\be
{\dot q}_i = H_{p_i} \;\;\; {\dot p}_i = -H_{q_i} \; (=L_{q_i}) \;\;\; .
\label{canleqnswithinCV2nd}
\ee
(A curve satisfying these equations is also called an extremal.) These are the 
Euler-Lagrange equations of a variational problem equivalent to the original 
one, in which both $q$s and $p$s are varied independently, namely
the problem
\be
\dd \int \left(\Sigma_i {\dot q}_ip_i - H(q_i,p_i,t)\right) \; dt \;\; = \;\; 0 
\;\; .
\label{canlvarnprobwithinCV2nd} 
\ee 
(For more details about eq. \ref{definepsubi} to \ref{canlvarnprobwithinCV2nd}, 
cf. e.g. Arnold (1989, Chap. 3.14, 9.45.C), Courant and Hilbert (1953, Chap. 
IV.9.3; 1962, Chap. I.6) and Lanczos (1986, Chap. VI.1-4).)

The requirement of a non-vanishing Hessian, eq. \ref{nonzerohessian} 
(equivalently: $\mid H_{{p}_i{p}_j}\mid \neq 0$), is a crucial assumption.  
Note in particular these two consequences.\\
\indent 1) The Hamiltonian cannot vanish identically. Proof: If we 
differentiate $H = \Sigma {\dot q}_ip_i - L = 0$ with respect to ${\dot q}_i$, 
we get $\Sigma_i \; L_{{\dot q}_i{\dot q}_j}{\dot q}_i = 0$; which contradicts 
eq. \ref{nonzerohessian}.\\
\indent 2) $L$ cannot be homogeneous of the first degree in the ${\dot q}_i$. 
That is, we cannot have: $L(q_i,\lambda{\dot q}_i,t) = \lambda L(q_i,{\dot 
q}_i,t)$. We shall see in Section \ref{sec:paraadditional} that this means the 
fundamental integral cannot be parameter-independent.

\subsection{Hypersurfaces and congruences}
\label{ssec;hypersurfs}
We consider a family of hypersurfaces in our region $G$ of $\mathR^{n+1}$
\be
S(q_i,t) = \sigma
\label{famyhyperintrodcd}
\ee
with $\sigma \in \mathR$ the parameter labelling the family, and $S$ a $C^2$ 
function (in all $n+1$ arguments). We assume this family covers the region $G$ 
{\em simply}, in the sense that through each point of $G$ there passes a unique 
hypersurface in the family. 

Let $C$ be a curve
\be
q_i = q_i(t)
\ee
of class $C^2$, that lies in $G$ and intersects each hypersurface in the family 
eq. \ref{famyhyperintrodcd} just once, but is nowhere tangent to a 
hypersurface. Then $\sigma$ is a function of $t$ along $C$, with
\be
\Delta :=\frac{d\sigma}{dt} = \Sigma_i \; \frac{\pl S}{\pl q_i}{\dot q}_i + 
\frac{\pl S}{\pl t}.
\label{defineDelta}
\ee  
By construction $\Delta \neq 0$. We will assume that the Lagrangian $L$ does 
not vanish along $C$. By a suitable labelling of the family of surfaces, we can 
secure
\be
\Delta > 0 \;\; {\rm  or } < 0 \;\; {\rm according \;\; as } \;\; L > 0 \;\; 
{\rm  or } < 0
\ee
for the line-element $(q_i,{\dot q}_i,t)$ of $C$. Then a tangential 
displacement along $C$ from $P:=(q_i,t)$ to $Q:=(q_i + dq_i,t +dt)$, i.e. a 
displacement with components $(dq_i,dt) = ({\dot q}_i,1)dt$, induces an 
increment $d\sigma$ in $\sigma$, and an increment $dI= L(q_i,{\dot q}_i,t)dt$ 
in $I = \int L \; dt$.

To connect this family of hypersurfaces with the calculus of variations, we now 
seek values of ${\dot q}_i$ at $P$ such that the direction at $P$ of the curve 
$C$, $({\dot q}_i,1)dt$, makes $dI/d\sigma$ a minimum with $d\sigma$ fixed. A 
necessary condition is that
\be
\frac{\pl }{\pl {\dot q}_i}\left(\frac{dI}{d\sigma} \right) = 0 , \;\;\; i = 
1,\dots,n.
\label{basicnecycondn}
\ee
But $\frac{dI}{d\sigma} = \frac{L}{\Delta}$ and $\Delta \neq 0$, so that eq. 
\ref{basicnecycondn} reads
\be
\frac{\pl L}{\pl{\dot q}_i} - \frac{L}{\Delta}\frac{\pl\Delta}{\pl{\dot q}_i} = 
0 \;\; ;
\label{2ndnecycondn}
\ee 
that is, using $\frac{\pl\Delta}{\pl{\dot q}_i} = \frac{\pl S}{\pl q_i}$ from 
eq. \ref{defineDelta},
\be
\frac{\pl L}{\pl{\dot q}_i} = \frac{L}{\Delta}\frac{\pl S}{\pl q_i} \;\; .
\label{3rdnecycondn}
\ee
A curve $C$, or its tangent vector $({\dot q}_i,1)$, that satisfies eq. 
\ref{3rdnecycondn}, is said to be in the direction of the {\em geodesic 
gradient determined by the family  of surfaces} \ref{famyhyperintrodcd}.

As it stands, this condition eq. \ref{3rdnecycondn} can at best yield minima of 
$dI/d\sigma$; while we are interested in minima of $dI/dt$. But there is a 
further condition on the family of surfaces eq.  \ref{famyhyperintrodcd} that 
implies that curves obeying eq. \ref{3rdnecycondn} are solutions of the 
variational problem; or rather, to be precise, extremals.

This condition has two equivalent forms; the first geometric in spirit, the 
second analytic. They are:\\
\indent (a): that the quantity $L/\Delta$ is constant on each surface, i.e. 
there is some real function $\phi$ such that 
\be
\frac{L}{\Delta} = \phi(\sigma)
\ee 
where we are to take the directional arguments in $L$ to refer to the geodesic 
gradient.\\ 
\indent (b): that $S$ solves the Hamilton-Jacobi  equation.

It is straightforward to show that (a) implies that we can re-parametrize the 
family of surfaces in such a way that $L = \Delta$ throughout the region $G$. 
That is to say: given (a), the family can be re-parametrized so that function 
$\phi$ is the constant function 1: $\phi(\sigma) = 1$. (Proof: any monotonic 
function $\psi$ gives a re-parametrization of the family, 
$\psi(S)=\psi(\sigma)$, with ${\bar \Delta}$ defined on analogy with $\Delta$ 
by ${\bar \Delta}:= \frac{d}{dt}{\psi(\sigma)} = \psi '(\sigma)\Delta$. 
Choosing $\psi(\sigma) := \int^{\sigma}_{\sigma_0} \phi(s) \; ds$ ($\sigma_0$ 
some constant) yields $\psi '(\sigma) = \phi(\sigma)$ so that $\frac{L}{{\bar 
\Delta}} \equiv \frac{L}{\psi '(\sigma)\Delta} \equiv \frac{\phi(\sigma)}{\psi 
'(\sigma)} = 1.$) 

So to show (a) and (b) equivalent, we will show that:\\
\indent (i) given (a) in this special form, i.e. given $L = \Delta$, $S$ solves 
the Hamilton-Jacobi equation; and conversely\\
\indent (ii) $S$ solving the Hamilton-Jacobi equation implies that 
$L=\Delta$.\\
But it will be clearest, before proving this equivalence, to present two  
consequences of $L= \Delta$, and introduce some terminology.

First: $L = \Delta$ implies that the geodesic gradient, eq. \ref{3rdnecycondn}, 
is now given by  
\be
\frac{\pl L}{\pl{\dot q}_i} = p_i = \frac{\pl S}{\pl q_i} \;\; .
\label{4thnecycondn}
\ee
where the first equation uses eq. \ref{definepsubi}.
Recall now our assumption that the determinant $\mid L_{{\dot q}_i{\dot 
q}_j}\mid \neq 0$, so that eq. \ref{definepsubi} can be solved in $G$ for the 
${\dot q}_i$ as functions of $q_i, p_i,t$: ${\dot q}_i = q_i(q_i,p_i,t)$. This 
now reads as
\be
{\dot q}_i = {\dot q}_i(q_i,\frac{\pl S}{\pl q_i},t) \; ,
\label{qdotifode}
\ee  
where the right-hand side is a function of $(q_i,t)$ alone (since $S$ is) and 
has continuous first order derivatives. Then the elementary existence theorem 
for solutions of first order ordinary differential equations implies that eq. 
\ref{qdotifode} defines an $n$-parameter family of curves in the region $G$, 
such that  each point in $G$ has a unique curve pass through it, and each curve  
is a solution of eq. \ref{qdotifode} in the sense that the components of its 
tangent vectors obey eq. \ref{qdotifode}. This family of curves is called the 
{\em congruence $K$ belonging to the family of surfaces} eq. 
\ref{famyhyperintrodcd}.

Second: $L = \Delta$ implies that the increment $dI$ in the fundamental 
integral $I = \int L \; dt$, in passing from a point $P_1$ on the surface 
$S(q_i,t) = \sigma_1$, to an adjacent surface $S = \sigma_1 + d\sigma$, along a 
curve of the congruence belonging to the family,  obeys 
\be
dI = \Delta dt = d \sigma \; .
\ee 
Integrating this result along members of the congruence, we get: the integral 
along a curve of the congruence, from any point $P_1$ on the surface $S(q_i,t) 
= \sigma_1$ to that point $P_2$ on the surface $S(q_i,t) = \sigma_2$ that lies 
on the same curve of the congruence, is the {\em same} for whatever point $P_1$ 
we choose. That is:
\be
\int^{P_2}_{P_1} L \; dt = \sigma_2 - \sigma_1 \; .
\label{samevalueforallexls}
\ee
Clearly, the converse also holds: if the fundamental integral taken along 
curves of the congruence has the same value for two hypersurfaces, however we 
choose the end-points $P_1,P_2$ lying in the hypersurfaces, then $L = \Delta$. 
So a family of surfaces satisfying the condition that $L = \Delta$ is called 
{\em geodesically equidistant} with respect to the Lagrangian $L$. (Courant and 
Hilbert (1962, Chap. II.9.2) say `geodetic', not `geodesic'; which has the 
advantage of avoiding `geodesic''s possibly confusing connotations of metric 
and-or connection.)

Carath\'{e}odory called a family of geodesically equidistant hypersurfaces, 
together with the congruence belonging to it, the {\em complete figure} (of the 
variational problem). As we shall see, the name is apt, since the complete 
figure is central to Hamilton-Jacobi theory. Also, the congruence is called 
{\em transversal} to the surfaces of the family. The analytical expression of 
transversality is that for a displacement $(\dd q_i,\dd t)$ tangential to a 
hypersurface in the family, $\dd S = 0$. That is: 
\be
\frac{\pl S}{\pl q_i}\dd q_i + \frac{\pl S}{\pl t}\dd t = 0 \;\; .
\label{transversyforS}
\ee

We turn to showing that: (i) $L = \Delta$ implies that $S$ solves the 
Hamilton-Jacobi equation.\\
\indent Proof: Eq. \ref{defineDelta} yields
\be
L(q_i,{\dot q}_i,t) = \Delta :=\frac{d\sigma}{dt} = \Sigma_i \; \frac{\pl 
S}{\pl q_i}{\dot q}_i + 
\frac{\pl S}{\pl t}
\ee
where ${\dot q}_i$ refers to the direction of the geodesic gradient, eq. 
\ref{qdotifode}, i.e. ${\dot q}_i = {\dot q}_i(q_i,\frac{\pl S}{\pl q_i},t)$. 
This yields 
\be
- \frac{\pl S}{\pl t} \; = \; -L(q_i,{\dot q}_i(q_i,\frac{\pl S}{\pl q_i},t),t) 
\; + \; \Sigma_i \; \frac{\pl S}{\pl q_i}\;{\dot q}_i(q_i,\frac{\pl S}{\pl 
q_i},t) \; .
\ee
But eq. \ref{legtrfmnsymmicform2nd} implies that the right-hand side is the 
Hamiltonian function, but with $p_i$ replaced by $\frac{\pl S}{\pl q_i}$ in 
accordance with eq. \ref{4thnecycondn}. Thus we have
\be
\frac{\pl S}{\pl t} + H(q_i,\frac{\pl S}{\pl q_i},t) = 0 \;\; ;
\label{hjRund}
\ee
which is the {\em Hamilton-Jacobi equation}.

This equation is also a sufficient condition of a family of surfaces being 
geodesically equidistant. That is, (ii): $S$ being a $C^2$ solution in $G$ of 
the Hamilton-Jacobi equation implies that $L=\Delta$, i.e. that the 
hypersurfaces of constant $S$ are geodesically equidistant.\\
\indent Proof: Given such a solution $S(q_i,t)$, let us define an assignment to 
each point of $G$ (sometimes called a {\em field}) by 
\be
p_i \equiv p_i(q_i,t) := \frac{\pl S}{\pl q_i} \; .
\label{pidefinedfromS}
\ee
By eq. \ref{nonzerohessian}, this determines a field ${\dot q}_i$ as in eq. 
\ref{qdotifode}, and hence a congruence. Then for the given solution $S$, a 
given member $C$ of the congruence, and two given parameter values 
$\sigma_1,\sigma_2$, we form the fundamental integral along $C$ between the 
points $P_1$ and $P_2$ where $C$ intersects the hypersurfaces corresponding to 
the parameter values $\sigma_1,\sigma_2$. Using  the Legendre transformation, 
eq. \ref{legtrfmnsymmicform2nd} and the fact that $S$ solves the 
Hamilton-Jacobi equation, eq. \ref{hjRund}, we obtain:
\be
\int^{P_2}_{P_1} L \; dt = - \int^{P_2}_{P_1} \left[H(q_i,\frac{\pl S}{\pl 
q_i},t) - \Sigma_i p_i{\dot q}_i \right] \; dt =
\int^{P_2}_{P_1} \left[\frac{\pl S}{\pl t}dt + \Sigma_i \frac{\pl S}{\pl 
q_i}{dq}_i \right] = \int^{P_2}_{P_1} \; dS =
\sigma_2 - \sigma_1.
\label{proveHJsufftgeodequidist}
\ee
 
To sum up: a family of hypersurfaces $S=\sigma$ is geodesically equidistant 
with respect to the Lagrangian $L$ iff $S$ is a solution of the Hamilton-Jacobi  
equation whose Hamiltonian $H$ corresponds by the Legendre transformation to 
$L$. And if this holds, the transversality condition, eq. \ref{transversyforS}, 
can be written (using eq. \ref{hjRund} and \ref{pidefinedfromS}) as
\be
p_i \dd q_i - H(q_i,p_i,t)\dd t = 0.
\footnote{Transversality can also be defined, without any use of a family of 
hypersurfaces, or even a function $S$, in terms of the fundamental integral 
being stationary as an end-point of the integral varies on a given single 
surface. Cf. e.g. Courant and Hilbert (1953, Chap. IV.5.2).}
\label{transversyforHp}
\ee 

\section{Canonical and Euler-Lagrange equations; fields of extremals 
}\label{sec:CELEfields}
We now study the properties of a congruence $K$ belonging to a family of 
geodesically equidistant surfaces. We first show that any curve of such a 
congruence obeys the canonical and Euler-Lagrange equations. Then we develop 
the ideas of: a field $q_i,p_i$ in the region $G$; and a field belonging to a 
family of (not necessarily geodesically equidistant) hypersurfaces. Finally we 
characterize those fields belonging to geodesically equidistant hypersurfaces. 

\subsection{Canonical and Euler-Lagrange equations}\label{ssec;CELE}
The family eq. \ref{famyhyperintrodcd} defines an assignment of $p_i := 
\frac{\pl S}{\pl q_i}$ to each point of a member $C$ of the congruence $K$. If 
we differentiate the definition of $p$ i.e. eq. \ref{pidefinedfromS} with 
respect to $t$ along $C$, and we differentiate the Hamilton-Jacobi  equation 
eq. \ref{hjRund}, and we then use the fact (from eq. \ref{canleqnswithinCV2nd}) 
that ${\dot q}_i = \frac{\pl H}{\pl p_i}$, we can eliminate the second 
derivatives of $S$ that arise in the differentiations, and get:
\be
{\dot p}_i = - \frac{\pl H}{\pl q_i} \; .
\ee
To this, we can adjoin ${\dot q}_i = \frac{\pl H}{\pl p_i}$, so as to get $2n$ 
first order ordinary differential equations obeyed by members of $K$
\be
{\dot p}_i = - \frac{\pl H}{\pl q_i} \;\;\; ; \;\;\;
{\dot q}_i = \frac{\pl H}{\pl p_i} \; \; .
\label{canldeddhyperss}
\ee

Note that according to this deduction, these two groups of equations have 
different statuses, despite their symmetric appearance. ${\dot p}_i = - 
\frac{\pl H}{\pl q_i}$ depends on $K$ belonging to a family of geodesically 
equidistant surfaces (i.e. on the Hamilton-Jacobi  equation ). But ${\dot q}_i 
= \frac{\pl H}{\pl p_i}$ are identities derived from the theory of the Legendre 
transformation (cf. eq. \ref{canleqnswithinCV2nd}). But this difference is not 
peculiar to our deduction's use of hypersurfaces. The same difference occurs in 
derivations of these equations in the calculus of variations with fixed 
end-points: in the most familiar case, in Lagrangian mechanics i.e. without use 
of the canonical integral; (cf. e.g. Lanczos (1986, p. 166-7).

From the canonical equations we can deduce the (Lagrangian form of the)  
Euler-Lagrange equations. We substitute $p_i = \frac{\pl L}{\pl {\dot q}_i}$ in 
the left-hand side, and $\frac{\pl H}{\pl q_i}= - \frac{\pl L}{\pl q_i}$ in the 
right-hand side, of the first of eq. \ref{canldeddhyperss}, to get
\be
\frac{d}{dt}\frac{\pl L}{\pl{\dot q}_i} - \frac{\pl L}{\pl q_i} = 0 \; .
\label{elfromhyperss}
\ee

\subsection{Fields}\label{ssec;fields}
To discuss fields, we need first to consider parametric representations of an 
arbitrary smooth congruence of curves covering our region $G$ simply. That is, 
we consider a congruence represented by $n$ equations giving $q_i$ as $C^2$ 
functions of $n$ parameters and $t$
\be
q_i = q_i(u_{\alpha},t) \;\;\; i = 1,\dots,n
\label{qfromcongruence}
\ee
where each set of $n$ $u_{\alpha} = (u_1,\dots,u_n)$ labels a unique curve in 
the congruence. Thus there is a one-to-one correspondence $(q_i,t) 
\leftrightarrow (u_{\alpha},t)$ in appropriate domains of the variables, with 
non-vanishing Jacobian
\be
\mid \frac{\pl q_i}{\pl u_{\alpha}} \mid \;\; \neq 0 \;\; .
\label{nonzeroJacobqalpha}
\ee

Such a congruence determines tangent vectors $({\dot q}_i,1)$ at each 
$(q_i,t)$; and thereby also values of the Lagrangian $L(q_i(u_{\alpha},t), 
{\dot q}_i(u_{\alpha},t),t)$ and of the momentum 
\be
p_i = p_i(u_{\alpha},t) = \frac{\pl L}{\pl {\dot q}_i} \;\; .
\label{pfromcongruence}
\ee
Conversely, a set of $2n \; C^2$ functions $q_i, p_i$ of $(u_{\alpha},t)$ as in 
eqs \ref{qfromcongruence} and \ref{pfromcongruence}, with the $q$s and $p$s 
related by $p_i = \frac{\pl L}{\pl {\dot q}_i}$, determines a set of tangent 
vectors, and so a congruence. Such a set of $2n$ functions is called a {\em 
field}; and if all the curves  of the congruence are extremals (i.e. solutions 
of the Euler-Lagrange equations), it is called a {\em field of extremals}.

We say a field {\em belongs to a} (not necessarily geodesically equidistant) 
{\em family of hypersurfaces} given by eq. \ref{famyhyperintrodcd}  iff 
throughout the region $G$ eq.s \ref{4thnecycondn} and \ref{qfromcongruence} are 
together satisfied, i.e. iff we have
\be
p_i = \frac{\pl }{\pl q_i}S(q_i,t) = \frac{\pl}{\pl q_i}S(q_i(u_{\alpha},t),t) 
\;\; .
\label{p=dSbydq;definebelonging}
\ee
One can show that a field belongs to a family of hypersurfaces iff for all 
indices $\alpha,\beta = 1,\dots,n$, the Lagrange brackets of the parameters of 
the field, i.e.
\be
[u_{\alpha},u_{\beta}] := \Sigma_i \left(\frac{\pl q_i}{\pl 
u_{\alpha}}\frac{\pl p_i}{\pl u_{\beta}} -
\frac{\pl q_i}{\pl u_{\beta}}\frac{\pl p_i}{\pl u_{\alpha}} \right)
\label{definelagebrackets}
\ee
vanish identically.\footnote{Cf. Rund (1966, p. 28-30). Warning: the role of 
Lagrange brackets in this theory is sometimes omitted even in excellent 
accounts, e.g. Courant and Hilbert (1962, Chap. II.9.4).}

We say that a field $q_i = q_i(u_{\alpha},t), p_i = p_i(u_{\alpha},t)$ is {\em 
canonical} if the $q_i, p_i$ satisfy eq. \ref{canldeddhyperss}. Now we will 
show that if a canonical field belongs to a family of hypersurfaces eq. 
\ref{famyhyperintrodcd}, then the members of the family are geodesically 
equidistant. 

Proof: Differentiating eq. \ref{p=dSbydq;definebelonging} with respect to $t$ 
along a member of the congruence, and substituting on the left-hand side  from 
the first of eq. \ref{canldeddhyperss}, we get
\be
- \frac{\pl H}{\pl q_i} = \frac{\pl^2 S }{{\pl q_i}{\pl q_j}}{\dot q}_j + 
\frac{\pl^2 S }{{\pl q_i}{\pl t}}
\ee
By the second of eq. \ref{canldeddhyperss}, this is
\be
\frac{\pl^2 S }{{\pl q_i}{\pl t}} + \frac{\pl H}{\pl q_i} + 
\frac{\pl H}{\pl p_j} \frac{\pl^2 S }{{\pl q_i}{\pl q_j}} = 0 
\ee
which is
\be 
\frac{\pl }{\pl q_i}\left(\frac{\pl S}{\pl t} + H (q_j,\frac{\pl S}{\pl q_j},t) 
\right) = 0
\ee
which is integrated immediately to give
\be
\frac{\pl S}{\pl t} + H(q_j,\frac{\pl S}{\pl q_j},t) = f(t)
\label{hjplusarbyfn}
\ee
with $f$ an arbitrary function of $t$ only. Now we argue (in the usual way, for 
the calculus of variations) that this function can be absorbed in $H$. For 
suppose the given Lagrangian were replaced by $\tilde{L}= L + f(t)$. The 
path-independence of the integral  $\int f(t) \; dt$ implies that $L$ and 
$\tilde{L}$ give equivalent variational problems, i.e. the same curves give 
stationary values for both $\int L \; dt$ and $\int {\tilde {L}} \; dt$. 
Besides, the definition of $p_i$, eq. \ref{definepsubi}, and the canonical 
equations eq. \ref{canleqnswithinCV2nd}  are unaffected, the only change in our 
formalism being that $H$ is replaced by $\tilde{H} = H - f(t)$. So assuming 
that $L$ is replaced by $\tilde{L}$ means that eq. \ref{hjplusarbyfn} reduces 
to the Hamilton-Jacobi equation, eq. \ref{hjRund}. The result now follows from 
result (ii) at the end of Section \ref{ssec;hypersurfs}.

This result is a kind of converse of our deduction of eq. 
\ref{canldeddhyperss}. We can sum up this situation by saying that the 
canonical  equations characterize any field belonging to a family of 
geodesically equidistant hypersurfaces. 

Finally, we should note an alternative to our order of exposition. We assumed 
at the outset a family of hypersurfaces, and then discussed an associated 
congruence and field. But one can instead begin with a single  arbitrary 
surface; then define the notion of an extremal being transverse to the surface 
(in terms of the fundamental integral being stationary as an end-point varies 
on the surface---cf. footnote 6); then define a field of such transverse 
extremals; and finally define other surfaces, geodesically equidistant to the 
given one, as surfaces $S =$ constant, where $S(q_i,t)$ is defined to be the 
value of the fundamental integral taken along a transverse extremal from the 
given surface ($S=0$) to the point $(q_i,t)$. This alternative order of 
exposition is adopted by Courant and Hilbert (1962, Chap. II.9.2-5), and (more 
briefly) by Born and Wolf (1999, Appendix I.2-4). It has the mild advantage 
over ours of clearly displaying the choice of an arbitrary initial surface; 
(which accords with the solution of a partial differential equation involving 
an arbitrary function just as the solution of an ordinary differential equation 
involves an arbitrary constant or constants). It will also come up again in 
Sections \ref{sec:fopdes} and \ref{sec:optics}.  

\section{Hilbert's independent integral}\label{sec:HbtIntgl}
A canonical field belonging to a geodesically equidistant family of 
hypersurfaces defines a line-integral which is independent of its path of 
integration. This integral, named after its discoverer Hilbert, is important 
not only in Hamilton-Jacobi theory, but also in aspects of the calculus of 
variations which we do not discuss, e.g. the study of conditions for the 
fundamental integral to take extreme values.

Suppose we are given a geodesically equidistant family  of hypersurfaces 
covering region $G$ simply. Consider two arbitrary points $P_1,P_2 \in G$ lying 
on hypersurfaces $S= \sigma_1, S=\sigma_2$ respectively; and consider an {\em 
arbitrary} $C^1$ curve $C:q_i=q_i(t)$ lying in $G$ and joining $P_1$ and $P_2$. 
We will write the components of the tangent  vector $(dq_i/dt,1)$ of $C$ as 
$(q'_i,1)$; for we continue to use the dot ${\dot {}}$ for differentiation 
along the geodesic gradient of the field belonging to $S$. Now consider the 
integral  along $C$ of $dS$, so that the integral  is trivially 
path-independent: 
\be
J := \int^{P_2}_{P_1} dS(q_i,t) = \sigma_2 - \sigma_1 = \int^{P_2}_{P_1}
\left(\frac{\pl S}{\pl q_i}q'_i + \frac{\pl S}{\pl t}\right) \; dt \;\; .
\label{Jtriviallypathindpdt}
\ee 
We can apply $p_i=\frac{\pl S}{\pl q_i}$ and the Hamilton-Jacobi  equation to 
the first and second terms of the integrand respectively, to get a 
path-independent integral 
\be
J = \int^{P_2}_{P_1} \left( p_i q'_i - H(q_i,p_i,t) \right) \; dt \;\; = \;\; 
\sigma_2 - \sigma_1 \;\; .
\label{Jcanlform}
\ee 
We can also Legendre transform to eliminate the $p_i$ in favour of ${\dot 
q}_i$, getting
\be
J = \int^{P_2}_{P_1}
\left(L(q_i,{\dot q}_i,t) + \frac{\pl L}{\pl {\dot q}_i}\left(q'_i - {\dot q}_i 
\right) \right) \; dt \;\; = \;\; \sigma_2 - \sigma_1 \;\; .
\label{Jlagnform}
\ee 
It is in this form that $J$ is usually called the {\em Hilbert integral}.

A field
\be
q_i = q_i(u_{\alpha},t) \;\;\; p_i = p_i(u_{\alpha},t)
\label{fieldforMayer}
\ee
(assumed to belong to a family of hypersurfaces in the sense of eq. 
\ref{p=dSbydq;definebelonging}) is called a {\em Mayer field} if substituting 
$q_i,p_i$ in the integral in eq. \ref{Jcanlform} yields an integral that is 
path-independent. So we have seen that a canonical field is a Mayer field. One 
can show that the converse holds, i.e. any Mayer field is canonical (Rund 1966, 
p. 33). So we have the result: a Mayer field is a canonical field belonging to 
a family of geodesically equidistant hypersurfaces. (It can also be shown that 
every extremal curve can be imbedded in a Mayer field.) 

Combining this with the results of Section \ref{sec:CELEfields}, we also have: 
the field eq. \ref{fieldforMayer} is a Mayer field iff the Lagrange brackets 
$[u_{\alpha},u_{\beta}]$ vanish and the field obeys the canonical equations eq. 
\ref{canldeddhyperss}.

\section{The parameter as an additional 
$q$-coordinate}\label{sec:paraadditional}
As we said at the start of Section \ref{ssec;cvreview}, our theory has depended 
from the outset on the choice of $t$; (cf. the fundamental integral eq. 
\ref{actionwithpara=t}). Indeed, we saw at the end of Section 
\ref{ssec;cvreview} that the non-vanishing Hessian eq. \ref{nonzerohessian} 
implies that 
$L$ cannot be homogeneous of the first degree in the ${\dot q}_i$; i.e. we 
cannot have for all $\lambda \in \mathR, L(q_i,\lambda{\dot q}_i,t) = \lambda 
L(q_i,{\dot q}_i,t)$. And we  shall shortly see that this implies that the 
fundamental integral cannot be parameter-independent. 

But for some aspects of the theory, especially the next Section's discussion of 
Hamilton-Jacobi theory as an integration theory for first order partial 
differential equations, it is both possible and useful to treat $t$ as a 
coordinate on a par with the $q$s. So in this Section, we describe such a 
treatment and the gain in symmetry it secures.

To have some consistency with our previous notation, we first consider a 
Lagrangian $L(q_{\alpha},{\dot q}_{\alpha},t)$ with $n-1$ coordinates 
$q_{\alpha}$, a parameter $t$ and derivatives ${\dot q}_{\alpha} = 
dq_{\alpha}/dt$. So note: in this Section, Greek indices run from 1 to $n-1$. 
So the fundamental integral along a curve $C:q_{\al} = q_{\al}(t)$ in a 
suitable region $G$ of $\mathR^n$ joining points $P_1,P_2$ with parameters 
$t_1,t_2$ is
\be
I = \int^{t_2}_{t_1} L(q_{\al},{\dot q}_{\al},t) \; dt \;\; .
\label{IfundlIntgl}
\ee
Now we introduce a real $C^1$ function $\tau(t)$ which is such that $d\tau/dt > 
0$ for all values of $t$ under consideration, but is otherwise arbitrary. We 
write derivatives with respect to $\tau$ using dashes, so that 
\be
q'_{\al} = {\dot q}_{\al}\left(\frac{dt}{d\tau} \right) \;\; .
\ee
So with $\tau_1 := \tau(t_1), \tau_2 := \tau(t_2)$, we can write eq. 
\ref{IfundlIntgl} as
\be
I = \int^{\tau_2}_{\tau_1} 
L\left(q_{\al},q'_{\al}\frac{d\tau}{dt},t\right)\frac{dt}{d\tau}d\tau \;\; .
\label{IfundlIntgl2ndform}
\ee
If we now write $q_n$ for $t$, so that we can write the coordinates on 
$\mathR^n$ as
\be
q_i = (q_{\al},t) = (q_{\al},q_n) \;\; i = 1,\dots,n {\rm{\;\; and \; write 
\;\;\;}} \frac{d t}{d \tau} = q'_n \neq 0 \;\;\; ,
\ee
then we can write eq. \ref{IfundlIntgl2ndform} as
\be
I = \int^{\tau_2}_{\tau_1} L^*(q_i,q'_i) \; d\tau
\label{Iintermsoftau}
\ee
where we have defined
\be
L^*(q_i,q'_i) := L^*(q_{\al},t,q'_{\al},q'_n) := 
L\left(q_{\al},\frac{q'_{\al}}{q'_n},t\right) \;\; .
\label{defineL*}
\ee

We stress that the values of the integrals eq.s \ref{IfundlIntgl} and 
\ref{Iintermsoftau} are equal. But the latter is by construction 
parameter-independent, since the choice of $\tau$ is essentially arbitrary. 
Also $L^*$ is by construction positively homogeneous of the first degree in the 
$q'_i = (q'_{\al},q'_n)$---i.e. for all positive numbers $\lambda, 
L^*(q_i,\lambda q'_i) = \lambda L(q_i, q'_i)$---irrespective of the form of the 
given Lagrangian $L$. In fact one can easily show that these two features are 
equivalent.

For the purposes of the next Section, we will now express the canonical  
equations of our variational problem, eq.s \ref{IfundlIntgl} or 
\ref{Iintermsoftau}, in the new notation. But  note: the total differentiation 
on the left-hand sides of the canonical  equations will still be 
differentiation with respect to the original parameter $t$---and so indicated 
by a dot. 

Writing the conjugate momenta of $L^*$ as $p^*$ for the moment, we have
\be
p^*_{\al} = \frac{\pl L^*}{\pl q'_{\al}} = \frac{\pl L}{\pl {\dot 
q}_{\al}}\frac{1}{q'_n}q'_n = p_{\al}
\ee  
so that these are identical with the original conjugate momenta; and so we will 
drop the $^*$ in $p^*_{\al}$. So the canonical equations for the indices 
$1,\dots,n-1$ are given, with the original Hamiltonian (Legendre) function 
$H(q_{\al},p_{\al},t)$ as defined in eq. \ref{legtrfmnsymmicform2nd}, by
\be
{\dot q}_{\al} = \frac{\pl H}{\pl p_{\al}} \;\; , \;\;\; 
{\dot p}_{\al} = - \frac{\pl H}{\pl q_{\al}} \;\; .
\label{canlfor1ton-1}
\ee
On the other hand, for the new $p_n$, we have
\be
p_n := \frac{\pl L^*}{\pl q'_n} \; = \; L \; - \; \Sigma_{\al} \; \frac{\pl 
L}{\pl {\dot q}_{\al}}
\frac{q'_{\al}}{q'_n} \; = \; L \; - \; \Sigma_{\al} \; p_{\al}{\dot q}_{\al}.
\ee
Comparing with the definition eq. \ref{legtrfmnsymmicform2nd} of the 
Hamiltonian (Legendre) function, this is
\be
p_n + H(q_{\al},p_{\al},t) = 0.
\label{pn=-H}
\ee
So differentiating $p_n$ with respect to the original parameter $t$ along an 
extremal gives 
\be
{\dot p}_n := \frac{dp}{dt} = -\frac{dH}{dt} = -\frac{\pl H}{\pl t} =
-\frac{\pl H}{\pl q_n}  
\ee
which fits well with eq. \ref{canlfor1ton-1}; (here $-\frac{dH}{dt} = 
-\frac{\pl H}{\pl t}$ follows as usual from the canonical equations, i.e. from 
$H$'s Poisson bracket with itself vanishing identically). But note that we also 
have ${\dot q}_n := \frac{dt}{dt} = 1 \neq \frac{\pl H}{\pl p_n} = -1$.

However, we can use the Hamilton-Jacobi equation to overcome this last 
``wrinkle''. i.e. to get a greater degree of symmetry. We can write the 
Hamilton-Jacobi equation of our variational problem eq. \ref{IfundlIntgl} as
\be
\Phi\left(q_i,\frac{\pl S}{\pl q_i}\right) =
H \left(q_{\al},\frac{\pl S}{\pl q_{\al}},q_n \right) + \frac{\pl S}{\pl q_n} = 
0 \; ,
\label{hjeqninPhi}
\ee
where $\Phi$ is defined as a function of $2n$ variables by
\be
\Phi(q_i,p_i) := H(q_{\al},p_{\al},q_n) + p_n \; .
\label{definePhi}
\ee
Now if the $p_{\al}$ in eq. \ref{definePhi} refer to a field of extremals 
belonging to a solution $S(q_{\al},q_n)$ of the Hamilton-Jacobi equation, so 
that $p_{\al} = \frac{\pl S}{\pl q_{\al}}$, then by eq.s \ref{pn=-H} and 
\ref{hjeqninPhi}, we also have: $p_n = \frac{\pl S}{\pl q_n}$. Besides, eq. 
\ref{definePhi} implies immediately
\be
\frac{\pl \Phi}{\pl q_i} = \frac{\pl H}{\pl q_i} \;\;\; , \;\;\;
\frac{\pl \Phi}{\pl p_{\al}} = \frac{\pl H}{\pl p_{\al}} \;\;\; , \;\;\;
\frac{\pl \Phi}{\pl p_n} = 1  \; (= {\dot q}_n \equiv \frac{dt}{dt})\;\;\; .
\label{immedfromdefinePhi}
\ee
It follows that we can write the canonical equations eq. \ref{canlfor1ton-1}, 
together with the relations for $q_n, p_n$, in a completely symmetrical way in 
terms of $\Phi$ as
\be
{\dot q}_i = \frac{\pl \Phi}{\pl p_i} \;\;\; , \;\;\;
{\dot p}_i = - \frac{\pl \Phi}{\pl q_i} \;\;\; ;
\ee
where, note again, the dot denotes differentiation with respect to $t$.

\section{Integrating first order partial differential 
equations}\label{sec:fopdes}
As mentioned in Section \ref{sec:intro}, we will {\em not} expound the usual 
approach (with Jacobi's theorem) to Hamilton-Jacobi theory as an integration 
theory for first order partial differential equations.\footnote{For an 
exposition cf. the references in footnote 4. As to the history: Whittaker 
(1959, p. 264,316) reports that this theory was developed by Pfaff and Cauchy 
in 1815-1819, using earlier results by Lagrange and Monge; i.e. well before 
Hamilton's and Jacobi's work!} Instead, we will in this Section briefly 
introduce another approach which exploits the results and concepts of the 
previous Sections; (for more details, cf. Rund, 1966, Chap. 2.8).

We will consider a partial differential equation of the form
\be
\Phi\left(q_i,\frac{\pl S}{\pl q_i}\right) = 0 \;\; , \;\; i = 1,\dots,n \; ; 
\;\; \mbox{ with }\;\; \frac{\pl \Phi}{\pl p_i} \neq 0 \;\; \mbox{ for at least 
one }i \;\; ,
\label{genlfopde}
\ee  
and $\Phi$ of class $C^2$ in all $2n$ arguments. One of the $i$ for which 
$\frac{\pl \Phi}{\pl p_i} \neq 0$ may be identified with $t$, but this is not 
necessary: as in the previous Section, our discussion can treat all coordinates 
of $\mathR^n$ on an equal footing. We shall also assume that (as suggested by 
the Hamilton-Jacobi equation) the unknown function $S$ does not occur 
explicitly in the equation; but this is not really a restriction, since one can 
show that the general case, i.e. an equation in which $S$ occurs, can be 
reduced to the form of eq. \ref{genlfopde} by introducing an additional 
independent variable.

So the initial value problem  we are to solve is: to find a function $S(q_i)$ 
($q_i \in G$) that satisfies eq. \ref{genlfopde} and that assumes prescribed 
values on a given $(n-1)$-dimensional $C^2$ surface, $V$ say, in $G$. We will 
indicate how to explicitly construct such a function by using a congruence of 
``canonical'' curves which solve a canonical system of ordinary differential 
equations; (so we reduce the integration of the partial differential equation 
to the problem of integrating ordinary differential equations). This canonical 
system of equations will be suggested by our previous discussion; and the 
strategy of the construction will be to adjust the congruence of curves from an 
initial rather arbitrary congruence, to one that provides a solution to eq. 
\ref{genlfopde}.\footnote{We remark at the outset that since---as in previous 
Sections---we work in a ``configuration space'', not its twice-dimensional 
``phase space'', there are many ``canonical congruences'', rather than a unique 
one; so that this sort of adjustment is possible.} 

Thus our previous discussion (especially Sections \ref{sec:CELEfields} and 
\ref{sec:paraadditional}) suggests we should consider the system of $2n$ 
ordinary differential equations, with a new parameter $s$
\be
{\dot q}_i := \frac{dq_i}{ds} = \frac{\pl \Phi(q_j,p_j)}{\pl p_i} \;\;\; , 
\;\;\;
{\dot p}_i := \frac{dp_i}{ds} = - \frac{\pl \Phi(q_j,p_j)}{\pl q_i} \;\;\; .
\label{chareqnsadjoined}
\ee
These are called the {\em characteristic equations} of eq. \ref{genlfopde}. A 
curve $q_i = q_i(s)$ of $\mathR^n$ that satisfies them is called a {\em 
characteristic curve} of eq. \ref{genlfopde}; it will be an extremal of a 
problem in the calculus of variations if eq. \ref{genlfopde} is the 
Hamilton-Jacobi equation of such a problem. Our approach to integrating eq. 
\ref{genlfopde} applies to these characteristic equations theorems about the 
existence and uniqueness of solutions of ordinary differential equations, so as 
to secure the existence and uniqueness of solutions to eq. \ref{genlfopde}.

Let us consider an $(n-1)$-parameter congruence of characteristic curves, with 
parameters $u_1,\dots,u_{n-1}$, so that we write 
\be
q_i = q_i(s,u_\al) \;\;\; , \;\;\; p_i = p_i(s,u_{\al}) \;\; .
\label{paramcongrceforintgn}
\ee
Since $\Phi$ is $C^2$, it follows from eq. \ref{chareqnsadjoined} that the 
functions \ref{paramcongrceforintgn} are $C^2$ in $s$. We will also assume that 
these functions are $C^2$ in the $u_{\al}$; and that this congruence covers the 
region $G$ simply, with
\be
\frac{\pl(q_1,q_2,\dots,q_n)}{\pl(s,u_1,\dots,u_{n-1})} \neq 0 \; ,
\label{nonzeroJacobforintgn}
\ee
so that we can invert the first set of eq. \ref{paramcongrceforintgn} for 
$(s,u_{\al})$, getting
\be
s = s(q_i) \;\; , \;\; u_{\al} = u_{\al}(q_i) \; .
\label{solveforsufnsofq}
\ee 
We shall also write (in $G$):
\be
\phi(s,u_{\al}) := \Phi(q_i(s,u_{\beta}),p_i(s,u_{\beta})) \; .
\label{phidefinedfromPhi}
\ee
One can now show:\\
\indent (i): $\phi$ of eq. \ref{phidefinedfromPhi} is an integral of the 
characteristic equations eq. \ref{chareqnsadjoined}, i.e. $\frac{d\phi}{ds} = 
0$;\\
\indent (ii): the Lagrange brackets $[u_{\al},s]$ and $[u_{\al},u_{\beta}]$ are 
constant along any member of the congruence defined by eq. 
\ref{chareqnsadjoined}.

We now make some assumptions about the relation of our characteristic 
congruence to the given surface $V$. We will assume that through each point of 
$V$ there passes a unique member of the congruence, and that the congruence is 
nowhere tangent to $V$. Thus each point in $V$ is assigned $n-1$ parameter 
values $u_{\al}$ and a value of $s$; so we can write $s$ on $V$ as a $C^2$ 
function of $u_{\al}$, the parameters of the unique curve through the point. 
Let us write this as $s = \sigma(u_{\al})$, so that the functions 
$a_I(u_{\al})$ defined by
\be
a_i(u_{\al}) : = q_i(\sigma(u_{\al}),u_{\al})
\label{defineai}
\ee 
are also $C^2$. Finally we will suppose that we seek a solution of eq. 
\ref{genlfopde} which takes the values $c(u_{\al})$ on $V$, $c$ prescribed 
$C^2$ functions. 

That completes the assumptions needed for the construction of a (local) 
solution of eq. \ref{genlfopde} (and the proof of its uniqueness). We end this 
Section by briefly describing the first steps of the construction. 

The theory of first order ordinary differential equations implies that the 
congruence of characteristic curves for eq. \ref{chareqnsadjoined} is 
determined if the values of $q_i$ and $p_i$ are prescribed on $V$. The initial 
values of $q_i$ are of course to be given by the $a_i$ of eq. \ref{defineai}. 
But as to the initial values of the $p_i$, i.e. $b_i$ defined by
\be
b_i(u_{\al}) : = p_i(\sigma(u_{\al}),u_{\al}) \;\; ,
\label{definebi}
\ee 
we have some choice. The strategy of the construction is, roughly speaking, to 
define a function $S$ on $G$, in such a way that when we adjust the $b_i$ so 
that $p_i = \frac{\pl S}{\pl q_i}$, $S$ becomes a solution of eq. 
\ref{genlfopde} in $G$, possessing the required properties.

We now define a function $z = z(s,u_{\al})$ on $G$ in terms of $V$, the  values 
$c(u_{\al})$ prescribed on $V$ and the given congruence; in effect, this $z$ 
will be the desired $S$, once the $b_i$ are suitably adjusted. For each point 
$P \in G$, with its $n$ parameter values $(s,u_{\al})$, the $s$-value of the 
intersection with $V$ of the unique curve through $P$  is given by $s = 
\sigma(u_{\al})$. We define the value of $z$ at $P$ by
\be
z(s,u_{\al}) := c(u_{\al}) + \int^s_{\tau = \sigma(u_{\al})} \;
\Sigma_i \; \left[p_i(\tau,u_{\al})\frac{\pl 
\Phi(q_i(\tau,u_{\al}),p_i(\tau,u_{\al}))}{\pl p_i}\right] \; d\tau \; \; ,
\label{definez}
\ee
where the integration is to be taken along the curve through $P$, from its 
point of intersection with $V$, to $P$. 

We will not go further into the construction of the desired $S$, except to make 
two remarks. (1): Note that eq. \ref{definez} implies in particular that 
$z(\sigma(u_{\al}),u_{\al}) = c(u_{\al})$.

(2): Differentiating eq. \ref{definez} with respect to $s$ and using the first 
set of eq. \ref{chareqnsadjoined} yields
\be
{\dot z} \equiv \frac{\pl z}{\pl s} = \Sigma_i \; p_i\frac{\pl \Phi}{\pl p_i} = 
\Sigma_i \; p_i{\dot q}_i \; \; .
\label{difftez}
\ee
This is analogous to the relation ${\dot S} = \Sigma_i p_i{\dot q}_i$ between a 
scalar function, such as a solution $S$ of the Hamilton-Jacobi equation and the 
field $q_i,p_i$ belonging to it, i.e. the field such that $p_i = \frac{\pl 
S}{\pl q_i}$; cf. eq. \ref{p=dSbydq;definebelonging}. Indeed, if we use eq. 
\ref{solveforsufnsofq} to define a function $S$ on $G$ by
\be
S(q_i) := z(s(q_i),u_{\al}(q_i))
\label{defineSintgnthy}
\ee 
then one can show (again, we omit the details!) that:\\
\indent (i) we can adjust the $b_i$ so as to make $p_i = \frac{\pl S}{\pl q_i}$ 
hold; and\\
\indent (ii) that this adjustment makes $S$, as defined by eq. 
\ref{defineSintgnthy} (and so \ref{definez}), a solution of eq. \ref{genlfopde} 
with the required properties. 

\section{The characteristic function and geometric optics}\label{sec:optics}
In this Section, we follow in Hamilton's (1833, 1834!) footsteps. We introduce 
the Hamilton-Jacobi equation via the characteristic function (as do most 
mechanics textbooks); and then apply these ideas to geometric optics---so our 
discussion will (at last!) make contact with physics. The main point will be 
that the correspondence in our formalism between canonical  extremals and 
geodesically equidistant hypersurfaces underpins the fact that both the 
corpuscular and wave conceptions of light can account for the phenomena, viz. 
reflection and refraction, described by geometric optics.\footnote{This is an 
example of what philosophers call ``under-determination of theory by data''. 
The escape from this sort of quandary is of course the consideration of other 
phenomena: in this case, the nineteenth-century study of diffraction and 
interference, which led to the rise of wave optics---cf. the start of Section 
\ref{sec:mechs}.}   

We assume that our region $G \subset \mathR^{n+1}$ is sufficiently small that 
between any two points $P_1 = (q_{1i},t_1), P_2 = (q_{2i},t_2)$ there is a 
unique extremal curve $C$. To avoid double subscripts, we will in this Section 
sometimes suppress the $i$, writing $P_1 = (q_1,t_1), P_2 = (q_2,t_2)$ etc. 
Then the value of the fundamental integral along $C$ is a well-defined function 
of the coordinates of the end-points; which we call the {\em characteristic 
function} and write as
\be
S(q_1,t_1; q_2,t_2) = \int^{t_2}_{t_1} L \; dt = 
\int^{t_2}_{t_1} (\Sigma_i p_i{\dot q}_i - H) \; dt =
\int \Sigma_i p_idq_i - Hdt  
\label{definecharfn}
\ee
where the integral is understood as taken along the unique extremal $C$ between 
the end-points, and we have used eq. \ref{legtrfmnsymmicform2nd}.

Making arbitrary small displacements $(\dd q_1,\dd t_1), (\dd q_2,\dd t_2)$ at 
$P_1,P_2$ respectively, and using the fact that the integral is taken along an 
extremal, we get for the variation in $S$
\begin{eqnarray}
\dd S := S(q_1 + \dd q_1,t_1 + \dd t_1; q_2 + \dd q_2,t_2 + \dd t_2) - 
S(q_1,t_1; q_2,t_2) = \nonumber \\
\frac{\pl S}{\pl t_1}\dd t_1 + \frac{\pl S}{\pl t_2}\dd t_2 + \Sigma_i \; 
\frac{\pl S}{\pl q_{1i}}\dd q_{1i} + \Sigma_i \;
\frac{\pl S}{\pl q_{2i}}\dd q_{2i} = 
\left[\Sigma_i \; p_i\dd q_i - H(q_j,p_j,t)\dd t \right]^{t_2}_{t_1} \;\; .
\label{varninS}
\end{eqnarray}
Since the displacements are independent, we can identify each of the 
coefficients on the two sides of the last equation in eq. \ref{varninS}, 
getting 
\begin{eqnarray}
\frac{\pl S}{\pl t_2} = -[H(q_i,p_i,t)]_{t=t_2} & , & 
\frac{\pl S}{\pl q_{2i}} = [p_i]_{t = t_2}
\label{identifycoefftslawvaryingaction;final} \\
\frac{\pl S}{\pl t_1} = [H(q_i,p_i,t)]_{t=t_1} & , & 
\frac{\pl S}{\pl q_{1i}} = - [p_i]_{t = t_1}
\label{identifycoefftslawvaryingaction;initial} 
\end{eqnarray} 
in which the $p_i$ refer to the extremal $C$ at $P_1$ and $P_2$.

These equations are remarkable, since they enable us, if we know the function  
$S(q_1,t_1,q_2,t_2)$ to determine all the extremals (in mechanical terms: all 
the possible motions of the system)---without solving any differential 
equations! For suppose we are given the initial conditions $(q_1,p_1,t_1)$, 
(i.e., in mechanical terms: the configuration and canonical momenta at time 
$t_1$), and also the function $S$. The $n$ equations $\frac{\pl S}{\pl q_1} = 
-p_1$ in eq. \ref{identifycoefftslawvaryingaction;initial} relate the $n+1$ 
quantities $(q_2,t_2)$ to the given constants $q_1,p_1,t_1$. So in principle, 
we can solve these equations by a purely algebraic process, to get $q_2$ as a 
function of $t_2$ and the constants $q_1,p_1,t_1$. Finally, we can get  $p_2$ 
from the $n$ equations $p_2 = \frac{\pl S}{\pl q_2}$ in eq. 
\ref{identifycoefftslawvaryingaction;final}. So indeed the extremals are found 
without performing integrations, i.e. just by differentiation and elimination: 
a very remarkable technique.\footnote{As Hamilton of course  realized. He 
writes, in the impersonal style of the day, that `Mr Hamilton's function $S$ 
... must not be confounded with that so beautifully conceived by Lagrange for 
the more simple and elegant expression of the known differential equations 
[i.e. $L$]. Lagrange's function {\em states}, Mr Hamilton's function would {\em 
solve} the problem. The one serves to form the differential equations of 
motion, the other would give their integrals' (1834, p. 514).} 

Substituting the second set of equations of eq.  
\ref{identifycoefftslawvaryingaction;final} in the first yields 
\be
\frac{\pl S}{\pl t_2} + H(q_2,\frac{\pl S}{\pl q_2}, t_2) = 0 \;\; .
\label{hjfinalfromvaryingaction}
\ee
So the characteristic function $S(q_1,t_1; q_2,t_2)$ considered as a function 
of the $n+1$ arguments $(q_2,t_2) = (q_{2i},t_2)$ (i.e. with $(q_1,t_1)$ fixed) 
satisfies the Hamilton-Jacobi equation.

Assuming that this solution $S$ is $C^2$, it follows from result (ii) of 
Section \ref{ssec;hypersurfs} that $S$ defines a family of geodesically 
equidistant hypersurfaces, namely the geodesic hyperspheres (for short: 
geodesic spheres) with centre $P_1 = (q_1,t_1)$. Thus the sphere with radius 
$R$ is given by the equation
\be
S(q_1,t_1;q_2,t_2) = R
\label{geodsphereR}
\ee
with $(q_1,t_1)$ considered fixed. So every point $P_2$ on this sphere is 
connected to the fixed centre $P_1 = (q_1,t_1)$ by a unique extremal along 
which the fundamental integral has value $R$. These extremals cut the spheres 
eq. \ref{geodsphereR} transversally.

These geodesic spheres about the various points $P_1$ are special families of 
hypersurfaces. For by taking envelopes of these spheres, we can build up 
successive members of an arbitrary family of geodesically equidistant 
hypersurfaces. This is the basic idea of {\em Huygens' principle} in geometric 
optics. Though Huygens first stated this idea as part of his wave theory of 
light, it can be stated entirely generally. Indeed, there is a rich theory 
here. We will not enter details\footnote{For details, cf. Baker and Copson 
(1950), Herzberger (1958). In optics, the Hamilton-Jacobi equation is often 
called the {\em eikonal} equation.}, but just state the main idea.

Thus consider some arbitrary solution $S(q_i,t)$ of the Hamilton-Jacobi  
equation
\be
\frac{\pl S}{\pl t} + H(q_i, \frac{\pl S}{\pl q_i},t) = 0
\label{hjinoptics}
\ee
and thereby the canonical field (congruence) $K$ belonging to it, for which 
$p_i = \frac{\pl S}{\pl q_i}$. Let $h_1, h_2$ be two hypersurfaces 
corresponding to values $\sigma_1,\sigma_2$ of $S$, i.e. $(q_i,t) \in h_j, (j = 
1,2)$ iff $S(q_i,t) = \sigma_j$. Let $P_1$ be in $h_1$, and let the canonical 
extremal $C$ through $P_1$ intersect $h_2$ in $P_2$. Then we already know from 
eq. \ref{samevalueforallexls} that the fundamental integral along $C$ is
\be
\int^{P_2}_{P_1} L \; dt = \sigma_2 - \sigma_1
\label{samevalueforallexls2nd}
\ee
so that $P_2$ is in the geodesic sphere centred on $P_1$ with radius $\sigma_2 
- \sigma_1$. Huygens' principle states that more is true: $h_2$ is the envelope 
of the set of geodesic spheres of radius $\sigma_2 - \sigma_1$ with centres on 
the hypersurface $h_1$.

As a final task for this Section, we briefly illustrate our formalism with 
another topic in geometric optics: namely, Fermat's ``least time'' principle, 
which states (roughly speaking) that a light ray between spatial points $P_1$ 
and $P_2$ travels by the path that makes stationary the time taken. This 
illustration has two motivations. First: together with the next Section's 
discussion, it will bring out the optico-mechanical analogy---and so prompt the 
transition to wave mechanics. 

Second: it illustrates how our formalism allows $t$ to be a coordinate like the 
$q_i$, even though it is singled out as the integration variable; (cf. Section 
\ref{sec:paraadditional}). In fact, there are subtleties here. For if one 
expresses Fermat's principle using time as the integration  variable, one is 
led to an integrand that is in general, e.g. for isotropic media, homogeneous 
of degree 1 in the velocities ${\dot q}_i$; and as noted in remark 2) at the 
end of Section \ref{ssec;cvreview}, this conflicts with our requirement of a 
non-vanishing Hessian (eq. \ref{nonzerohessian}), i.e. with our construction of 
a canonical formalism. So illustrating our formalism with Fermat's principle in 
fact depends on using a spatial coordinate as integration variable (parameter 
along the light's path). As we will see in a moment, this gives an integrand 
which is in general, even for isotropic media, not homogeneous of degree 1 in 
the velocities---so that we can apply the theory of Sections \ref{sec:CVtoHJ} 
onwards. 

So now our preferred coordinate $t$ will be (not time, as it will be in 
mechanics) but one of just three spatial coordinates $(q_1,q_2,t)$ for ordinary 
Euclidean space. In fact, applications of geometric optics, e.g. to optical 
instruments which typically have an axis of symmetry, often suggest a natural 
choice of the coordinate $t$. 

At a point $P = (q_1,q_2,t)$ in an optical medium, a direction is given by 
direction ratios $({\dot q}_1,{\dot q}_2,{\dot t}) = ({\dot q}_1,{\dot 
q}_2,1)$. (So note: the subscripts 1 and 2 now refer to the first and second of 
three spatial axes ``at a single time''---and not to initial and final 
configurations.) The speed of a ray of light through $P$ in this direction will 
in general depend on both position and direction, i.e. on the five variables 
$(q_i,{\dot q}_i,t), i = 1,2$; and so the speed is denoted by $v(q_i,{\dot 
q}_i,t)$. If $c$ is the speed of light {\em in vacuo}, the {\em refractive 
index} is defined by
\be
n(q_i,{\dot q}_i,t):= c/v(q_i,{\dot q}_i,t) \; .
\label{definen}
\ee
If $n$ is independent of the directional arguments ${\dot q}_i$ (respectively: 
positional arguments $q_i,t$), the medium is called {\em isotropic} 
(respectively: {\em homogeneous}).

Now let the curve $C:q_i = q_i(t)$ represent the path of a light-ray between 
two points $P_1,P_2$ with parameter values $t =t_1, t =t_2$. Then the time 
taken to traverse this curve (the {\em optical length} of the curve) is
\be
T = \int^{t_2}_{t_1} \frac{ds}{v} = 
\int^{t_2}_{t_1} \frac{n(q_i,{\dot q}_i,t)}{c} \; [({\dot q}_1)^2 + ({\dot 
q}_2)^2 + 1]^{\frac{1}{2}} \; dt =  \int^{t_2}_{t_1} L \; dt \;\; ,
\label{opticallength}
\ee 
where we have defined
\be
L(q_i,{\dot q}_i,t) := \frac{n(q_i,{\dot q}_i,t)}{c} \; [({\dot q}_1)^2 + 
({\dot q}_2)^2 + 1]^{\frac{1}{2}} \;\; .
\label{defineopticalL}
\ee
However, our discussion will not be concerned with this special form of $L$. We 
will only require that $L$ be $C^2$, and that the Hessian does not vanish, i.e. 
eq. \ref{nonzerohessian} holds. One immediately verifies that this is so for 
isotropic media; (in fact the Hessian is $\frac{n(q_i,t)^2}{c^2}[({\dot q}_1)^2 
+ ({\dot q}_2)^2 + 1]^{-2} \neq 0$).

We can now connect our discussion with the principles of Fermat and Huygens. We 
can again take Fermat's principle in the rough form above, viz. that a light 
ray between points $P_1$ and $P_2$ travels by the path that makes stationary 
the time taken. It follows that if light is instantaneously emitted from a 
point-source located at $P_1 = (q_{1i},t_1)$ (where now we revert to using `1' 
to indicate an initial location), then after a time $T$ the light will register 
on a surface, $F(T)$ say, such that each point $P_2 = (q_{2i},t_2)$ on $F(T)$ 
(where  similarly, `2' indicates a final location) is joined to $P_1$ by an 
extremal along which the fundamental integral assumes the common value $T$. 
This surface is the {\em wave-front} for time $T$, due to the point-source 
emission from $P_1$. Clearly, the family of wave-fronts, as $T$ varies, is 
precisely the family of geodesic spheres (for $L$ as in eq. 
\ref{defineopticalL}) around $P_1$.

Using the Hamilton-Jacobi  equation 
eq. \ref{hjinoptics} (now with just three independent variables $q_1,q_2,t$), 
we can readily generalize this, so as to describe the construction of 
successive wave-fronts, given an initial wave-front. Given an arbitrary 
solution $S(q_1,q_2,t)$ of eq. \ref{hjinoptics}, and an initial hypersurface 
$h_1$ given by $S(q_i,t) = \sigma_1$, we can construct at each point $P_1 \in 
h_1$ the unique extremal of the canonical field belonging to the family of 
hypersurfaces of constant $S$. By Fermat's principle, each such extremal can 
represent a ray emitted from $P_1$. If we define along each such extremal the 
point $P_2$ such that that fundamental integral $\int^{P_2}_{P_1} L \; dt$ 
attains the value $T$, then the locus of these points $P_2$ is the surface 
$S=\sigma_1 + T$. 
Thus we construct a family of geodesically equidistant 
hypersurfaces.\footnote{The vector $p_i = \frac{\pl S}{\pl q_i}$ is longer the 
more rapidly $S$ increases over space, i.e. the more rapidly the light's time 
of flight increases over space. So Hamilton called $p_i$ the {\em vector of 
normal slowness.}} To sum up: each solution of the Hamilton-Jacobi equation 
represents a family of wave-fronts, and the canonical field belonging to a 
family represents the corresponding light rays.

\section{From the optico-mechanical analogy to wave mechanics}\label{sec:mechs}
The rise of wave optics in the nineteeth century led to geometric optics being 
regarded as the short-wavelength regime of a wave theory of light. So its 
equations and principles, such as the Hamilton-Jacobi equation and Fermat's and 
Huygens' principles, came to be seen as effective statements derived in the 
short-wavelength limit of the full wave theory. But the details of these 
derivations are irrelevant here.\footnote{cf. e.g. Born and Wolf (1999, Chap. 
3.1, 8.3.1); Taylor (1996, Section 6.6-6.7) is a brief but advanced 
mathematical discussion.}

For us the relevant point is that (as is often remarked: e.g. Synge (1954, 
Preface), Rund (1966, p. 100)): once one considers this development, together 
with the optico-mechanical analogy as stated so far (i.e. as it stood for 
Hamilton), it is natural to speculate that there might be a wave mechanics, 
just as there is a wave optics. That is, it is natural to speculate that 
classical mechanics might describe the short-wavelength regime of a wave 
mechanics, just as geometric optics describes the short-wavelength regime of a 
wave optics. This is of course precisely what deBroglie, and then 
Schr\"{o}dinger, did. To be more specific, using our Hamilton-Jacobi 
perspective: they proposed that $S$ represented, not an ensemble of systems 
each fully described by its classical mechanical state $(q,p)$, but a property 
of an individual system.\footnote{Of course, successful proposals often seem 
``natural'' in hindsight; and some authors (e.g. Goldstein (1950, p. 314)) 
maintain that deBroglie's and Schr\"{o}dinger's proposal would have seemed 
merely idle speculation if it had  been made independently of the introduction 
of Planck's constant and the subsequent struggles of the old quantum theory. 
Indeed, even in that context it was obviously: (i) daringly imaginative 
(witness the fact that the $S$ wave propagates in multi-dimensional 
configuration space); and (ii) confusing (witness the interpretative struggles 
over the reality of the wave-function). In any case, whether the proposal was 
natural or not---after all, `natural' is a vague word---all can now agree that 
their achievement was enormous.}

In this Section, we give a simple sketch of this proposal. But we shall not 
give details of deBroglie's and Schr\"{o}dinger's own arguments, which are 
subtle and complicated: (Dugas (1988, Part V, Chap. 4) gives some of this 
history). Our sketch is formal, though in the textbook tradition (Rund (1966, 
pp. 99-109) and Goldstein (1950, pp. 307-314)); (various books give fuller 
accounts e.g. using the concepts of Fourier analysis and the group velocity of 
a wave-packet, e.g. Messiah (1966, pp. 50-64), Gasiorowicz (1974, pp. 27-32)). 
More precisely: we will first  describe how when we apply Hamilton-Jacobi 
theory to a classical mechanical system, the $S$-function defines for each time 
$t$ surfaces of constant $S$ in configuration space, so that by varying $t$ we 
can calculate the velocity with which these ``wave-fronts'' propagate (in 
configuration space). So far, so classical. But then we will postulate that 
these wave-fronts are surfaces of constant phase of a time-dependent  
complex-valued wave-function on configuration space. This will lead us, with 
some heuristic steps, to the Schr\"{o}dinger equation and so to wave mechanics.
   
Let us consider a classical mechanical system with holonomic ideal constraints, 
on which the constraints are solved so as to give a $n$-dimensional 
configuration space $Q$, on which the $q_i$ are independent variables. More 
technically, $Q$ is a manifold, on which the $q_i$ are a coordinate  system, 
and on which the kinetic energy defines a metric. But we shall not go into this 
aspect: we shall simply assume $Q$ is equipped with the usual Euclidean metric 
on $\mathR^n$, and that the $q_i$ are rectangular coordinates. We further 
assume that any constraints are time-independent (scleronomous); i.e. any 
configuration  in $Q$ is possible for the system throughout the time period in 
question. The result of these assumptions is that the region $G \subset 
\mathR^{n+1}$ for which the formalism of Sections \ref{sec:CVtoHJ} has been 
developed is now assumed to be an `event space' or `extended configuration 
space' of the form $Q \times T$, where $T \subset \mathR$ is some real interval 
representing a period of time. Finally, we will assume that our system is 
conservative, with energy $E$.

Now we will presume, without rehearsing the usual equations (cf. especially 
Section \ref{ssec;cvreview} and eq.s \ref{definecharfn} to 
\ref{hjfinalfromvaryingaction}), that using the above assumptions, the 
Lagrangian and Hamiltonian mechanics of our system has been set up. So if 
$S(q_i,t) = \sigma$ is a family of geodesically equidistant hypersurfaces 
associated with the system (each hypersurface $n$-dimensional), the family 
covering our region $G$ simply, then $S$ satisfies the Hamilton-Jacobi equation 
in the form $\frac{\pl S}{\pl t} + E = 0$. This can be immediately integrated 
to give, for some function $S^*$ of $q_i$ only,
\be
S(q_i,t) = S^*(q_i) - Et \; .
\label{introS*}
\ee
(So the $p_i$ of the canonical  field depend only on $S^*$: $p_i := \frac{\pl 
S}{\pl q_i} = \frac{\pl S^*}{\pl q_i}$.) So the hypersurfaces of our family can 
be written as 
\be
S^*(q_i) = Et + \sigma \; .
\label{famyintermsofS*}
\ee
For any fixed $t$, a hypersurface of constant $S$, considered as a hypersurface 
in the configuration space $Q$ (a hypersurface  of dimension $n-1$, i.e. 
co-dimension 1), e.g. the surface $S(q_i,t) = \sigma_1$, coincides with a 
hypersurface of constant $S^*$: for this example, the surface $S^* = \sigma_1 + 
Et$. But while the surfaces of constant $S^*$ are time-independent, the 
surfaces of constant $S$ vary with time. So we can think of the surfaces of 
constant $S$ as propagating through $Q$. With this picture in mind, let us 
calculate their velocity.

(We can state the idea of surfaces in $Q$ of constant $S$ more rigorously, 
using our assumption that the region $G \subset \mathR^{n+1}$ is of the form $Q 
\times T$. This implies that any equation of constant time, $t =$ constant, 
defines a $n$-dimensional submanifold of $G$ which is a ``copy'' of $Q$; let us 
call it $Q_t$. Each hypersurface in eq. \ref{famyintermsofS*} defines a 
$(n-1)$-dimensional submanifold of $Q_t$ (a hypersurface in $Q_t$ of 
co-dimension 1) given by 
\be
S^*(q_i) = Et + \mbox{ constant, (with $t$ = constant)}.
\label{surfaceinRn}
\ee
Then, as in the previous paragraph: fixing the constant $\sigma$ but letting 
$t$ vary, and identifying the different copies $Q_t$ of $Q$, we get  a family 
of $(n-1)$-dimensional submanifolds of $Q$, parametrized by $t$. This can be 
regarded as a wave-front propagating over time through the configuration space 
$Q$.)  

Let us fix a constant $\sigma$ and a time $t$; let $P=(q_i) \in Q$ be a point 
on the surface $S = S^* - Et = \sigma$; and consider the normal to this surface 
(pointing in the direction of propagation) at $P$. (So the $i$th component 
$n_i$ of the unit normal is $
n_i = \mid \nabla S^* \mid^{-1}\frac{\pl S^*}{\pl q_i} \; $.) Consider a point 
$P' = (q_i + dq_i)$ that lies a distance $ds$ from $P$ along this normal: (so 
$dq_i = n_i ds$). $P'$ is on a subsequent wave-front (i.e. with the same value 
$\sigma$ of $S$, but not of $S^*$) at time $t+dt$, where by eq. \ref 
{famyintermsofS*}
\be
dS^* \; = \; \Sigma_i \; \frac{\pl S^*}{\pl q_i}dq_i \; = \; E \; dt ;
\label{fixdt}
\ee
which, dividing by $ds$, yields
\be
\frac{dS^*}{ds} \; \equiv \; \mid \nabla S^* \mid \; = \; \Sigma_i \; \frac{\pl 
S^*}{\pl q_i}\frac{dq_i}{ds} \; = \; E \; \frac{dt}{ds} .
\label{fixdtbyds}
\ee
But we also have
\be
p_i = \frac{\pl S^*}{\pl q_i} \;\; \Rightarrow \;\; \; p := \; \mid p \mid \; = 
\; \mid \nabla S^* \mid \; .
\label{p=gradS}
\ee
Combining these equations, eq. \ref{fixdtbyds} and \ref{p=gradS}, we deduce 
that the speed $u$ of  the wavefront $S= \sigma$, i.e. $u := \frac{ds}{dt}$, is  
\be
u = \frac{E}{p} \; .
\label{u}
\ee

So far, so classical. But now we postulate that the wave-fronts eq. 
\ref{famyintermsofS*} (or \ref{surfaceinRn}) are surfaces of constant phase of 
a suitable time-dependent complex-valued function $\psi$ on $Q$. This 
postulate, together with some heuristic steps (including a judicious 
identification of Planck's constant!), will give us a heuristic derivation of 
the Schr\"{o}dinger equation. We will assume to begin with that we can write 
the postulated function $\psi = \psi(q_i,t)$ as
\be
\psi = R(q_i,t) \exp[-2\pi i(\nu t - \phi(q_i))] \;\; ,
\label{intropsi}
\ee
with $R$ and $\phi$ real; so that $\nu t - \phi$ is the phase, and (apart from 
$R$'s possible $t$-dependence) $\nu$ is the frequency associated with $\psi$. 
Then our postulate is that there is some constant $h$ such that 
\be
h(\nu t - \phi(q_i)) = (Et - S^*(q_i)) \;\; .
\label{constphasecondn}
\ee
But this must hold for all $q_i,t$, so that
\be
E = h \nu \;\;\; ; \;\;\; S^*(q_i) = h\phi(q_i) \;\; .
\label{basicproportions}
\ee
So the postulated frequency is proportional to the system's energy. Then, using 
our previous calculation of the speed $u$, and the relation  $u = \lambda \nu$ 
with $\lambda$ the wavelength, we deduce that the wavelength is inversely 
proportional to the magnitude of the system's momentum. That is:   
\be
u = \lambda \nu = \frac{E}{p} \;\;\; \Rightarrow \;\;\; \lambda = \frac{h}{p} 
\;\; .
\label{deducelamdahp}
\ee
Substituting eq. \ref{basicproportions} in eq. \ref{intropsi}, we can write 
$\psi$ as
\be
\psi = R(q_i,t) \exp \left[\frac{2\pi i}{h}(S^*(q_i) - Et) \right] =
R(q_i,t) \exp \left[\frac{i}{\hbar}(S^*(q_i) - Et) \right]
\label{psiwithSandE}
\ee
where we have defined $\hbar := \frac{h}{2 \pi}$.

Assuming now that $R$ has no $q_i$-dependence, differentiation of eq. 
\ref{psiwithSandE} with respect to $q_i$ yields 
\be
\frac{\pl \psi}{\pl q_i} = \frac{i}{\hbar}\frac{\pl S^*}{\pl q_i}\psi 
\label{difftepsiqi}
\ee
Recalling that $p_i = \frac{\pl S^*}{\pl q_i}$, this is an eigenvalue equation, 
and suggests that we associate with the $i$th component of momentum $p_i$ of a 
system whose $R$ has no $q_i$-dependence, the operator ${\hat p}_i$ on 
wave-functions $\psi$ defined by  
\be
 {\hat p}_i := \frac{\hbar}{i}\frac{\pl }{\pl q_i} \; , \; i = 1,\dots, n \; .
\label{definemommopor}
\ee
Let us postulate this association also for $q_i$-dependent $R$. Then this  
suggests we also associate with the energy of the system, the operator ${\hat 
H}$ on wave-functions defined by  
\be
 {\hat H} := H(q_i, {\hat p}_i,t) \; \; ,
\label{definehamnopor}
\ee
(where we understand $q_i$, and functions of it, as operating on wave-functions 
by ordinary multiplication). 

But assuming now that $R$ has no $t$-dependence,  differentiation of eq. 
\ref{psiwithSandE} with respect to $t$ yields 
\be
i\hbar\frac{\pl \psi}{\pl t} = E\psi
\label{difftepsit}
\ee
suggesting we should associate with the energy of a system, the operator ${\hat 
E}$ on wave-functions defined by
\be
{\hat E} := i\hbar \frac{\pl }{\pl t}
\label{defineenergyopor}
\ee
(By the way, this definition is also motivated by treating time as a coordinate 
along with the $q_i$; cf. the discussion in Section \ref{sec:paraadditional}. 
Thus eq. \ref{definemommopor} suggests we define ${\hat p}_{n+1} := 
\frac{\hbar}{i}\frac{\pl }{\pl t}$; and when this is combined with eq. 
\ref{defineenergyopor}, we get
\be
{\hat p}_{n+1} + {\hat E} = 0
\label{pn+1-energyopor}
\ee
which is analogous to eq. \ref{pn=-H}.)

If for general $R(q_i,t)$ we endorse both these suggestions---i.e. we identify 
the actions on eq. \ref{psiwithSandE} of these two suggested operators , eq. 
\ref{definehamnopor} and \ref{defineenergyopor}---then we get 
\be
 {\hat H}\psi = i\hbar \frac{\pl \psi}{\pl t} \;\; ;
\label{Schrodeqn}
\ee
which, once we identify $h$ as Planck's constant, is the Schr\"{o}dinger 
equation.

\section{A glance at the pilot-wave theory}\label{sec:pwthy}
So much by way of sketching the Hamilton-Jacobi perspective on the heuristic 
route to wave mechanics. In this final Section, I will briefly return to this 
volume's question `{\em Quo vadis}, quantum mechanics?', i.e. to the 
foundations of quantum theory. First, I want to stress that Hamilton-Jacobi 
theory remains an important ingredient in various research programmes in this 
field. Prominent among these is the trajectory representation of quantum 
mechanics, pioneered by Floyd, and Faraggi and Matone.  I cannot go into 
details, but would recommend, as places to begin reading, both Floyd (2002) and 
Faraggi and Matone (2000). (Besides, Section 1 of the latter ends with some 
references to other research programmes that use Hamilton-Jacobi theory.)

 I shall instead end on Hamilton-Jacobi theory in the context of another 
prominent research programme (related to the trajectory representation): 
deBroglie's and Bohm's pilot-wave theory. Again, this is a large topic, and we 
only wish to advertise the recent work of Holland (2001, 2001a).

First, we recall (Bohm (1952, p. 169), Bohm and Hiley (1993, p. 28), Holland 
(1993, pp. 69,134)) that:\\
\indent (i): Writing $\psi = R(q_i,t) \exp(iS(q_i,t)/\hbar)$ ($R, S$ real) in 
the one-particle Schr\"{o}dinger equation, eq. \ref{Schrodeqn} with ${\hat H} 
:= \frac{\hbar^2}{2m}\nabla^2 + V$ gives
\be
\frac{\pl S}{\pl t} + \frac{1}{2m}(\nabla S)^2 + Q + V = 0 \;\; \mbox{with} 
\;\; Q := \frac{-\hbar^2}{2m}\frac{\nabla^2 R}{R} \; ,
\label{qm'hj'eqn}
\ee
which looks like the classical Hamilton-Jacobi equation (cf. eq. \ref{hjRund}) 
of a particle in an external potential that is the sum of $V$ and $Q$, which 
Bohm called the `quantum potential'; indeed Bohm and Hiley call eq. 
\ref{qm'hj'eqn} the ``quantum Hamilton-Jacobi equation''; and
\be
\frac{\pl \rho}{\pl t} + \frac{1}{m}\nabla \cdot (\rho \nabla S) = 0 \;\; 
\mbox{with} \;\; \rho := R^2 \;\; .
\label{contyeqn}
\ee
\indent (ii): These equations suggest the quantum system comprises both a wave, 
propagating according to the Schr\"{o}dinger equation, and a  particle, which 
has (a) a continuous trajectory governed by the wave according to the  guidance 
equation
\be
m\frac{d q_i}{dt} = \frac{\pl S}{\pl q_i}\mid_{q_i = q_i(t)} \;\; ;
\label{guidance}
\ee
and (b) a probability distribution given at all times by $\mid \psi \mid^2 = 
R^2$.

Besides, comments and equations similar to those in (i) and (ii) apply when we 
insert $\psi = R \exp(iS/\hbar)$ into the many-particle Schr\"{o}dinger 
equation (Bohm (1952, p. 174), Bohm and Hiley (1993, p. 56 et seq.), Holland 
(1993, pp. 277 et seq.)).

So far, so good. But Holland (2001, p. 1044) points out that the relation of 
pilot-wave theory to classical Hamilton-Jacobi theory is {\em not} transparent. 
In particular, he points out:\\  
\indent 1): The guidance law eq. \ref{guidance} is `something of an enigma'. It 
looks like one half of a canonical transformation that trivializes the motion 
of a classical system (by transforming to a set of phase space coordinates that 
are constant in time. But what about the other half; and more generally, can 
eq. \ref{guidance} be somehow related to a Hamiltonian or Hamilton-Jacobi 
theory?\\
\indent 2): $Q$'s dependence on $S$ (through eq. \ref{contyeqn}) means that the 
``quantum Hamilton-Jacobi equation'' eq. \ref{qm'hj'eqn} in effect contains 
higher derivatives of $S$---wholly unlike a classical Hamilton-Jacobi equation.

So Holland undertakes an extensive investigation of this relation. More 
precisely, he undertakes to formulate the pilot-wave theory as a Hamiltonian 
theory. He does this by assessing a treatment of $Q$ as a field function of 
$q_i$ on a par with the classical potential $V$; i.e. a treatment that takes as 
the Hamiltonian of the (one-particle) system
\be
H(q_i,p_i,t) = \frac{1}{2m}\Sigma_i p^2_i + Q(q_i,t) + V(q_i,t)
\label{HforQasfnofq}
\ee
He emphasises that such a treatment faces three obstacles. In brief, they 
are:\\ 
\indent (a): As we just mentioned in 2), $Q$ depends on $S$ and so presumably, 
by $p = \frac{\pl S}{\pl q}$, on $p$. So in a Hamiltonian (phase space) 
treatment, it seems wrong to take $Q$ as a function only of $q$.\\
\indent (b): The free choice of initial positions and momenta in a Hamiltonian 
treatment will mean that most motions, projected on $q$, do not give the 
orthodox quantal distribution, in the way that eq. \ref{guidance} and $\mid 
\psi \mid^2 = R^2$ does.\\
\indent (c): Is such a treatment compatible with the Hamiltonian description of 
the Schr\"{o}dinger equation? For it to be so, we have to somehow formulate the 
particle-wave interaction so as to prevent a back-reaction on the wave.

However, Holland goes on to show (2001, 2001a) that these obstacles can be 
overcome. That is, he vindicates the proposal, eq. \ref{HforQasfnofq}, with a 
Hamiltonian theory of the interacting wave-particle system. But we cannot enter 
details. It must suffice to list some features of his work. In short, his 
approach:\\ 
\indent (1): generalizes a canonical treatment of a classical particle and 
associated ensemble;\\ 
\indent (2): necessitates the introduction of an additional field of which the 
particle is the source;\\
\indent (3): makes the ``quantum Hamilton-Jacobi  equation'' and the continuity 
equation eq. \ref{contyeqn} (and other equations for the evolution of particle 
and field variables) come out as Hamilton equations;\\
\indent (4): interprets $p = \frac{\pl S}{\pl q}$ as a constraint on the phase 
space coordinates of the wave-particle system; \\
\indent (5): gives a general formula expressing the condition that the 
particle's phase space distribution, projected on $q$, gives the orthodox 
quantal distribution; and finally,\\
\indent (6): yields a Hamilton-Jacobi theory of the wave-particle system.

To conclude: I hope to have shown that Hamilton-Jacobi theory, understood from 
the perspective of the calculus of variations, gives us  insight into both 
mechanics and optics---and that, as illustrated by this last Section, 
Hamilton-Jacobi theory is an important ingredient in current attempts to answer 
the question, `{\em Quo vadis}, quantum mechanics?'.

{\em Acknowledgements}:---\\
It is a pleasure to thank: Avshalom Elitzur and Nancy Kolenda for a superb 
meeting; the participants, other audiences, and Alexander Afriat, Robert 
Bishop, Michael Hall, and especially Ned Floyd and Peter Holland, for 
conversations and comments on a previous version.

\section{References}\label{sec:refs}
V. Arnold (1989), {\em Mathematical Methods of Classical Mechanics}, 
Springer-Verlag (second edition).\\
B. Baker and E. Copson (1950), {\em The Mathematical Theory of Huygens' 
Principle}, Oxford University Press.\\
S. Benton (1977), {\em The Hamilton-Jacobi Equation: a Global Approach}, 
Academic Press.\\
D. Bohm (1952), `A suggested interpretation of the quantum theory in terms of 
``hidden'' variables, I', {\em Physical Review} {\bf 85}, pp. 166-180.\\
D. Bohm and B. Hiley (1993), {\em The Undivided Universe}, Routledge.\\
M. Born and E. Wolf (1999), {\em Principles of Optics}, Cambridge University 
Press (7th edition).\\
J. Buchwald (1989), {\em The Rise of the Wave Theory of Light}, University of 
Chicago Press.\\
J. Butterfield (2003), `David Lewis Meets Hamilton and Jacobi',  forthcoming in 
a supplementary issue of  {\em Philosophy of Science}.\\
J. Butterfield (2003a), `Some Aspects of Modality in Analytical mechanics', 
forthcoming in {\em Formal Teleology and Causality}, ed. M. St\"{o}ltzner, P. 
Weingartner, Paderborn: Mentis.  \\
R. Courant and D. Hilbert (1953), {\em Methods of Mathematical Physics}, volume 
I, Wiley-Interscience (Wiley Classics 1989).\\
R. Courant and D. Hilbert (1962), {\em Methods of Mathematical Physics}, volume 
II, Wiley-Interscience (Wiley Classics 1989).\\ 
R. Dugas (1988), {\em A History of Mechanics}, Dover; reprint of a 1955 French 
original.\\
A. Faraggi and M. Matone (2000), `The Equivalence Postulate of Quantum 
Mechanics', {\em International Journal of Modern Physics} {\bf A15}, pp. 
1869-2017; and available at hep-th/9809127. \\
E. Floyd (2002), `The Philosophy of the Trajectory Representation of Quantum 
Mechanics', in {\em Gravitation and Cosmology: From the Hubble Radius to the 
Planck Scale}, Proceedings of a Symposium in Honor of J.P.Vigier's 80th 
Birthday, Kluwer; and available at quant-ph/0009070.\\
S. Gasiorowicz (1974), {\em Quantum Physics}, John Wiley.\\
H. Goldstein (1950), {\em Classical Mechanics}, Addison-Wesley.\\
W. Hamilton (1833), `On a General Method of Expressing the Paths of Light, and 
of the Planets, by the Coefficients of a Characteristic Function', {\em Dublin 
University Review}, October 1833, 795-826.\\
W. Hamilton (1834), `On the Application to Dynamics of a General Mathematical 
Method previously Applied to Optics', {\em British Association Report}, 1834, 
513-518.\\
M. Herzberger (1958), {\em Modern Geometrical Optics}, Interscience 
Publishers.\\
P. Holland (1993), {\em The Quantum Theory of Motion}, Cambridge University 
Press.\\
P. Holland (2001), `Hamiltonian Theory of Wave and Particle in Quantum 
Mechanics Part I', {\em Nuovo Cimento} {\bf 116B} 1043-1069.\\
P. Holland (2001a),`Hamiltonian Theory of Wave and Particle in Quantum 
Mechanics Part II', {\em Nuovo Cimento} {\bf 116B} 1143-1172.\\
F. John (1971), {\em Partial Differential Equations}, Springer Verlag.\\
M. Kline (1970), {\em Mathematical Thought from Ancient to Modern Times}, 
Oxford University Press.\\
C. Lanczos (1986), {\em The Variational Principles of Mechanics}, Dover (4th 
edition).\\
R. Littlejohn (1992), `The Van Vleck Formula, Maslov Theory and Phase Space 
Geometry', {\em Journal of Statistical Physics} {\bf 68}, 7-50.\\
J. Logan (1994), {\em An Introduction to Non-linear Partial Differential 
Equations}, John Wiley.\\
A. Messiah (1966), {\em Quantum Mechanics}, vol. I, North Holland: John 
Wiley.\\
H. Rund (1966), {\em The Hamilton-Jacobi Theory in the Calculus of Variations}, 
Van Nostrand.\\
I. Stakgold (1967), {\em Boundary Value Problems of Mathematical Physics}, 
Macmillan.\\
J. Synge (1954), {\em Geometric Mechanics and deBroglie Waves}, Cambridge 
University Press.\\
M. Taylor (1996), {\em Partial Differential Equations}, volume 1, Springer 
Verlag.\\
A. Webster (1950), {\em Partial Differential Equations of Mathematical Physics} 
(ed. S.Plimpton), Hafner.\\
E. Whittaker (1952), {\em A History of the Theories of Aether and Electricity}, 
volume 1, Nelson.\\
E. Whittaker (1959), {\em A Treatise on the Analytical Dynamics of Particles 
and Rigid Bodies}, Cambridge University Press (4th edition).

\end{document}